\documentclass{my-aims}

\usepackage{subfigure}

\usepackage{txfonts}
\def\typeofarticle{Research article}

\def\ppages{Published in: AIMS Mathematics 4 (2019), no.~1, 86--98}
\def\DOI{10.3934/Math.2019.1.86}
\def\Received{01 November 2018}
\def\Accepted{21 January 2019}
\def\Published{xx January 2019}
\usepackage{hyperref}
\makeatletter
\def\UrlAlphabet{%
	\do\a\do\b\do\c\do\d\do\e\do\f\do\g\do\h\do\i\do\j%
	\do\k\do\l\do\m\do\n\do\o\do\p\do\q\do\r\do\s\do\t%
	\do\u\do\v\do\w\do\x\do\y\do\z\do\A\do\B\do\C\do\D%
	\do\E\do\F\do\G\do\H\do\I\do\J\do\K\do\L\do\M\do\N%
	\do\O\do\P\do\Q\do\R\do\S\do\T\do\U\do\V\do\W\do\X%
	\do\Y\do\Z}
\def\UrlDigits{\do\1\do\2\do\3\do\4\do\5\do\6\do\7\do\8\do\9\do\0}
\g@addto@macro{\UrlBreaks}{\UrlOrds}
\g@addto@macro{\UrlBreaks}{\UrlAlphabet}
\g@addto@macro{\UrlBreaks}{\UrlDigits}
\makeatother

\numberwithin{equation}{section}

\makeatletter
\renewcommand{\@biblabel}[1]{#1\hfill \hspace{-0.2cm}}
\makeatother

\theoremstyle{plain}

\theoremstyle{remark}

\theoremstyle{definition}

\emergencystretch=\maxdimen
\begin{document}

\title{The spread of a financial virus through Europe and beyond}


\author{Olena Kostylenko$^1$,
Helena Sofia Rodrigues$^{1, 2}$ 
and
Delfim F. M. Torres$^1$\corrauth}

\shortauthors{the Author(s)}


\address{$^1$Center for Research and Development in Mathematics and Applications (CIDMA),\\
Department of Mathematics, University of Aveiro, 3810-193 Aveiro, Portugal\\[0.3cm]
$^2$School of Business Studies, 
Polytechnic Institute of Viana do Castelo, 
4930-678 Valen\c{c}a, Portugal}

\corraddr{Email: delfim@ua.pt; Tel: +351234370668; Fax: +351234370066.}


\begin{abstract}
We analyse the importance of international 
relations between countries on the financial stability.
The contagion effect in the network is tested by implementing 
an epidemiological model, comprising a number of European countries 
and using bilateral data on foreign claims between them. Banking statistics 
of consolidated foreign claims on ultimate risk bases, obtained from the Banks 
of International Settlements, allow us to measure the exposure of contagion 
spreading from a particular country to the other national banking systems. 
We show that the financial system of some countries, experiencing the debt crisis, 
is a source of global systemic risk because they threaten the stability 
of a larger system, being a global threat to the intoxication of the world economy 
and resulting in what we call a ``financial virus''. Illustrative simulations 
were done in the NetLogo multi-agent programmable modelling environment
and in MATLAB.
\end{abstract}


\maketitle


\section{Introduction}

\noindent The global crisis of 2008 had the most devastating consequences in the world economy \cite{imf}. 
One of the main causes for the beginning and aggravation of the crisis was the strengthening 
of international economic interdependence. Primarily, the crisis hit the financial system 
and the debtor countries. As a result, systemic financial risks occurred, 
and the crisis spread to other countries.

The global financial system is a kind of configuration of numerous interrelations 
between national economies, and every day the world economy becomes more like 
a unified space with a network nature. Failure of one subject of the financial system 
generates a chain reaction through interconnections and causes shocks and systemic risk. 
This risk is associated with the incapability of one of the participants to perform 
their obligations (or to accomplish them properly), which leads to the interruption 
in the functioning of other participants and, thus, of the entire system. 
The Bank for International Settlements (BIS) provides the following definition 
of systemic financial risks: ``the risk that the failure of a participant to meet 
its contractual obligations may in turn cause other participants to default, 
with the chain reaction leading to broader financial difficulties'' \cite{bisreport}. 
Therefore, the systemic risk is the likelihood of negative changes in the financial 
system and the economy of a particular country that affect the financial stability 
of the global market \cite{investopedia,Kaufman}.

Crises continue to occur at different economic levels, both at micro and macro 
levels \cite{kosnaz}, which make the economy an interesting object of study. 
The interest of scientists to this field is increasing after some collapse in the economy.
The contribution of scientists in the study of the world economy is huge and has increased 
rapidly over the past decade. In \cite{inci}, the authors investigated contagion between 
international equity markets using the local correlation. The contagion effects among 
the stock markets were investigated in \cite{alexakis} using the asymmetric dynamic 
conditional correlation dynamics. Authors in \cite{giudici} investigated Corporate 
Default Swap spreads using the vector autoregressive regression with correlation 
networks in their model. Also, one of the interesting types of research in this area 
is the work \cite{ahnert}, where the authors studied information contagion due to 
the counterparty risk and examined its effects on banks \emph{ex-ante} choices and systemic risk.

Mathematical epidemiology is widely developed, as described in \cite{brauer}, 
and has wide application in various fields of science \cite{rodrigues2,ozturka,cannarella,rodrigues,wu}. 
However, the use of epidemiological models in the economy is scanty and the economy 
has not been studied yet completely. Thus, the economy needs to be investigated in order 
to prevent possible negative consequences, since the systemic risks accumulate in 
the world financial system and become a general threat to the new global crisis \cite{rockinger}.
The study of the systemic financial risks allows to characterize comprehensively 
the current picture of the global financial world, and also to develop new methods 
of protection against global threats. The significance of global systemic financial 
risks is increased by their complexity in the identification, estimation, and developing 
methods for their calculation and minimization \cite{cerutti}.

A key feature of global systemic financial risks is the potential 
infection of the world economy with a financial virus \cite{paltalidis}. 
For example, if some European Union countries are a source of global systemic risk, 
as they experience a debt crisis, then they threaten the stability 
of a larger system, which is a global threat.
For this reason, it is necessary to study the spread of financial 
viruses in the world economic network. The complex study of country 
interrelations shows which national banking systems are most exposed 
to a particular country, both on an immediate counterparty 
basis and on an ultimate risk basis \cite{cerutti}.
Our research focuses on total foreign claims on an ultimate risk basis, 
which captures lending to a borrower in any country that is guaranteed 
by an entity that resides in the counterparty country. The object of study 
is the process of infection spreading through network interconnections. 
Moreover, we investigate economic relations between the subjects of the 
global financial system, which arise in the process of managing systemic 
financial risks. The aim is to study the process of spreading the infection 
through network interconnections, identify regularities, and whenever 
possible give recommendations for minimization risks in global scale management.

The scientific novelty of our study consists in modelling and investigating 
the process of contagion in the network using epidemiological models. 
The research was done with statistical data from BIS \cite{bis} 
on the volumes of consolidated foreign claims on ultimate risk bases 
in a number of countries, and data of countries credit rating 
from the Guardian Datablog \cite{guardata}.

The paper is structured as follows. 
In Section~\ref{sec:02} of ``Methodology'', 
the basic concepts of network and epidemiological models 
are introduced as well as the data used for the considered models.
The results of modelling and various scenarios 
of contagion spreading are presented in Section~\ref{sec:03}
of ``Results''. We end our work
with Section~\ref{sec:04} of ``Conclusions''.


\section{Methodology}
\label{sec:02}

\noindent Based on the network nature of the global economy, described above, 
the systemic risk can be considered as a network risk, which causes infection of networks.

Our method for investigating the spread of a virus 
in the financial system consists of six steps:
$1)$ to build the network; 
$2)$ to define the virus transmitting rate and recovery rate;
$3)$ to visualise the process of virus transmission in the network 
by implementing a multi-agent programmable modelling environment 
in NetLogo \cite{netlogo};
$4)$ to run the spreading process in a closed population by solving 
the Kermack--McKendrick SIR model \cite{kermack} in the multi-paradigm 
numerical computing environment MATLAB \cite{matlab};
$5)$ to compare results between dynamics of infection in the network 
and dynamics obtained by solving the SIR system of differential equations;
$6)$ to confirm or disprove the economic reasonableness of the results.


\subsection{Network} 

\noindent Network analysis is well used in various fields of science \cite{bartlett}: 
in computer science, to describe the internet topology \cite{alderson}; 
in social sciences, to describe the evolution and spread of ideas 
and innovations in societies \cite{hufnagel}; in ecology, 
to model networks of ecological interactions \cite{rezende}; 
in biology, to investigate the neurovascular structure 
of the human brain \cite{david}; in biochemistry, to infer 
how selection acts on metabolic pathways \cite{proulx,albert}; 
as well as in economics, to study financial contagion 
in the banking system \cite{paltalidis,garas}.

Many mechanisms and quantitative tools for describing networks 
have been provided by research in graph theory. 
Networks are mathematically described as graphs. 
There are different types of graphs: random graphs, 
small-world graphs, scale-free graphs, and others.

A network consists of multiple nodes connected to each other.
In this research we construct a fully connected network, which includes $n$ nodes. 
This network is also known as a complete graph, denoted by $K_{n}$. 
The complete graph is a regular graph, where each vertex has the same 
degree $n-1$, and $K_{n}$ always has $n(n-1)/2$ links. 
It means that all nodes are interconnected, i.e., 
each vertex has the same number of neighbours. 

In graph theory, a finite graph is often represented as an adjacency matrix: 
\begin{equation}
\label{amatrx}
A = 
\left[
\begin{array}{cccc}
a_{11}& a_{12} &\ldots & a_{1n}\\
a_{21}& a_{22} &\ldots & a_{2n}\\
\vdots& \vdots &\ddots & \vdots\\
a_{n1}& a_{n2} &\ldots & a_{nn}
\end{array}\right],
\end{equation}
where elements $a_{ij}$ equal to zero or one, 
respectively for disconnected and connected vertices.
Such matrix is the basis for network building.
A network construction provides a good visualization of its structure, 
knowledge and understanding, which allows us to compute 
the epidemic dynamics and to predict a spreading phenomena.


\subsection{Epidemiological model}

\noindent The epidemic spreading can be described by many models. 
Epidemiological models, in their majority, are based on dividing 
the population according to the disease status of their individuals.
The main models describe the proportion of population that is infected, 
susceptible to infection, and recovered after a disease \cite{MyID:417}.

In our study, we use the classical Kermack--McKendrick SIR model \cite{kermack}, 
which considers such factors as infection spreading and recovery \cite{kosroddelf}:
\begin{eqnarray}
\label{eqSIR}
\begin{cases}
\displaystyle \frac{dS(t)}{dt} = - \beta S(t) I(t),\\[0.3cm]
\displaystyle \frac{dI(t)}{dt} = \beta S(t) I(t) - \gamma I(t),\\[0.3cm]
\displaystyle \frac{dR(t)}{dt} = \gamma I(t),
\end{cases} 
\end{eqnarray}
$t \in [0,T]$, subject to the initial conditions
\begin{equation}
\label{eqIC}
S(0) = S_0, \quad I(0) = I_0, \quad R(0) = R_0.
\end{equation}
The SIR model \eqref{eqSIR}--\eqref{eqIC} expresses the spread among 
the population compartments as a system of differential equations, 
where $S$, $I$ and $R$ refer to the number of susceptible, infectious 
and recovered individuals, respectively, in a constant population 
of $N$ individuals for all time $t$:
\begin{equation}
\label{eqN}
S(t)+I(t)+R(t) = N, \quad t\in [0,T], \quad T > 0. 
\end{equation}
System \eqref{eqSIR} describes the relationship between the three compartments:
a susceptible individual changes its state to infected with probability $\beta$ 
(the contagion spreading rate), while an infected changes its state to recovered 
with probability $\gamma$ (the speed of recovery). 
These parameters are assumed constant for the entire sample.


\subsection{Data}

\begin{figure}[ht!]
\centering
\includegraphics[scale=0.25]{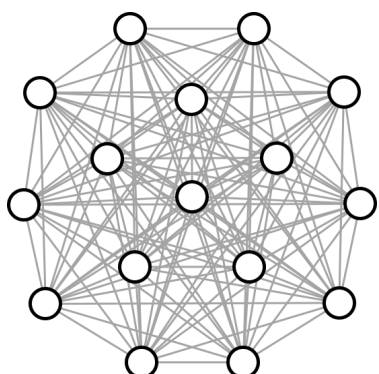} 
\caption{Fully connected network, where each country is represented as a node
and edges indicate the existence of a link between countries.} 
\label{fnetwork}
\end{figure}

\noindent Sixteen European and Non-European developed countries 
were chosen based on statistical data from the Bank for International 
Settlements (BIS) for the end of the year $2012$ \cite{bis}.
The number of countries $N = 16$ is fixed throughout the contamination time.
They connected to the network (see Figure~\ref{fnetwork}) according 
to the adjacency matrix \eqref{amatrx} as follows:
\begin{equation}
\label{matrx}
A = 
\left[
\begin{array}{cccc}
0& 1 &\ldots & 1\\
1& 0 &\ldots & 1\\
\vdots& \vdots &\ddots & \vdots\\
1& 1 &\ldots & 0
\end{array}\right],
\end{equation}
where the elements $a_{ij} $ are equal to one for connected vertices, 
and zero for disconnected. The connection is provided by the presence 
of bilateral foreign claims on an ultimate risks basis. 
The diagonal elements are all zero, since loops are not determined by 
statistical data of amounts outstanding from BIS \cite{bis}.

In our work, we assume that only one country is contagious at the initial time.
Thus, the values of initial conditions \eqref{eqIC} for the SIR model are as follows: 
$S(0) = 15$, $I(0) = 1$, and $R(0) = 0$. 

We also assume that the initially infected country $I$ cannot 
fulfil all of its obligations to other countries (for example, by domestic reasons). 
This means that all foreign claims $\alpha_{ij}$ of a counterparty country $i$ are infected. 
A contribution of infected debts to the total amount of claims from all countries, 
defines the value of $\beta$ parameter:
\begin{equation}
\label{eqbeta}
\beta_{i} = \frac{\sum\limits_{j=1}^{16} \alpha_{ij}} 
{\sum\limits_{i=1}^{16}\sum\limits_{j=1}^{16} \alpha_{ij}},
\quad i \in \left\{1, \ldots, 16\right\}. 
\end{equation}
The values of the infection spreading rate $\beta_i \in [0, 1]$ and $\sum_{i=1}^{16} \beta_{i}=1$ or $100\%$.
Thus, the more outstanding debts in the total amount of debts, imply the higher possibility of infection. 
Statistical information was taken from the BIS consolidated international banking statistics 
on an ultimate risk basis \cite{bis}. It is the most appropriate source for measuring 
the aggregate exposures of a banking system to a given country \cite{avdjiev}.

The recovery rate was calculated according to country's credit rating:
\begin{equation}
\label{eqgamma}
\gamma_{i} = \frac{1} {101-C_{i}}, 
\quad i \in \left\{1, \ldots, 16\right\}. 
\end{equation}
Here, $\gamma_i$ implies that it takes $\frac{1}{\gamma_i} = 101-C_{i}$ time steps to recover.
The credit rating $C_{i}$ takes into account not only countries' debt, 
but also assets. This measures the ability to fulfil their obligations as borrowers 
-- the probability of recovery. The data of countries' credit rating is taken from 
the Guardian Datablog \cite{guardata} and converted by ourselves to the numerical 
representation based on the rating table from \cite{TE}, where the credit rating 
is shown by country's credit worthiness between $100$ (riskless) and $0$ (likely to default).
Thus, a susceptible country $S$ can obtain contagion if it has a relationship 
with an infected country $I$, and if it has not enough money in reserve 
to cover possible risk losses.
 
\begin{figure}[ht!]
\centering
\includegraphics[scale=0.53]{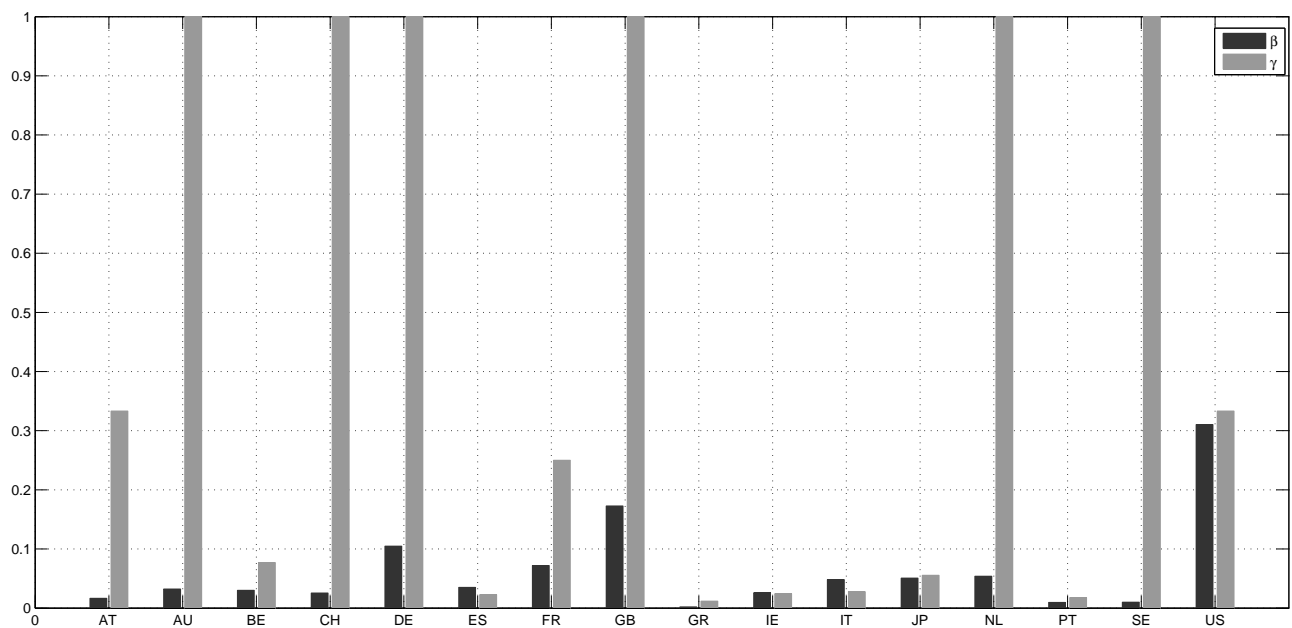}
\vspace*{0.3cm}
\caption{Summary statistics of $\gamma$ and $\beta$ parameters for 
the 16 European and Non-European developed countries considered in our study.} 
\label{fdata}
\end{figure}

The values of contagion spreading rate and the speed 
of recovery are given in Figure~\ref{fdata}.


\section{Results}
\label{sec:03}

\noindent We now present the obtained results.
For comparison, all countries are grouped according to the value 
of the recovery parameter $\gamma$ (see Table~\ref{tabular:gamma}).
\begin{table}[ht!]
\doublerulesep 0.1pt
\tabcolsep 7.8mm
\centering
\caption{\rm Grouping of countries depending on $\gamma$ parameter.}
\label{tabular:gamma}
\vspace*{2mm}
\renewcommand{\arraystretch}{1.3}
\setlength{\tabcolsep}{22pt}
\begin{center}
\footnotesize{
\begin{tabular*}{12.5cm}{ccc} \hline\hline\hline
\raisebox{-2ex}[0pt][0pt]{Group 1: $\gamma \leq 0.1$} &
\raisebox{-2ex}[0pt][0pt]{Group 2:	$0.1 < \gamma \leq 0.5$} &
\raisebox{-2ex}[0pt][0pt]{Group 3: $0.5 < \gamma \leq 1$} \\ \\ \hline
BE & AT & AU \\
ES & FR & CH \\
GR & US	& DE \\
IE &    & GB \\
IT &    & NL \\
JP &    & SE \\
PT &    &	 \\ \hline\hline\hline
\end{tabular*}}
\end{center}
\renewcommand{\arraystretch}{1}
\end{table}
We compare the results depending on the belonging of the initially 
infected country to a particular group. For illustrative purposes, we consider
the dynamics of the chain propagation reaction for three cases: when the contagion 
process starts 1) from Portugal (PT, Group~1); 2) from United States 
(US, Group~2); and 3) from Switzerland (CH, Group~3).
Both network and SIR model simulation results are consistent.


\subsection{Network}

\noindent To investigate the dynamics of infection spread in the network, 
we use the NetLogo agent-based programming language and integrated 
modelling environment \cite{netlogo}. It is well-recognized that its
visualization makes it easy to understand chain reaction processes 
\cite{netlogprog}. 

\begin{figure}[hp]
\centering
\begin{subfigure}
[$T_{PT}=0$]{\label{fnetpt0}
\includegraphics[width=0.23\textwidth]{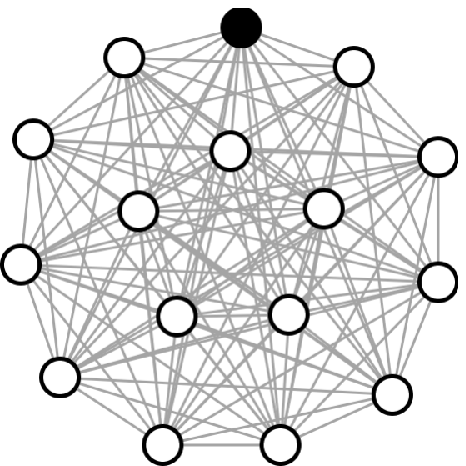}}
\end{subfigure}
\hfill
\begin{subfigure}[$T_{PT}=9$]{\label{fnetpt9}
\includegraphics[width=0.23\textwidth]{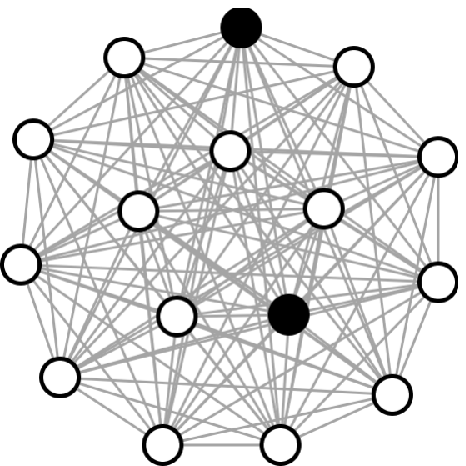}}
\end{subfigure}
\hfill
\begin{subfigure}[$T_{PT}=15$]{\label{fnetpt15}
\includegraphics[width=0.23\textwidth]{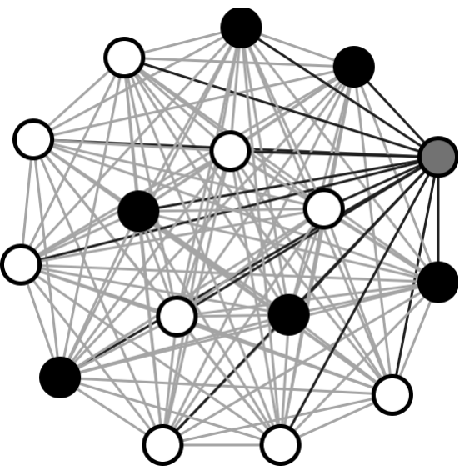}}
\end{subfigure}
\hfill
\begin{subfigure}[$T_{PT}=48$]{\label{fnetpt48}
\includegraphics[width=0.23\textwidth]{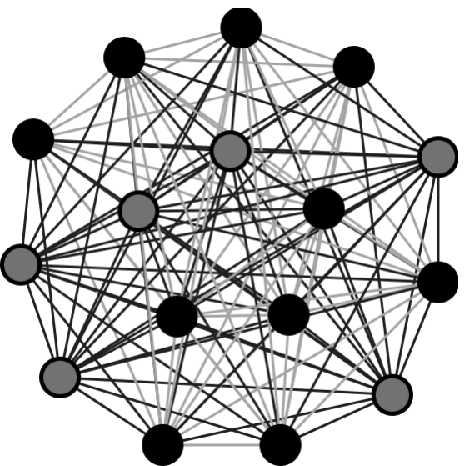}}
\end{subfigure}
\vfill
\begin{subfigure}[$T_{PT}=280$ ]{\label{fnetpt280}
\includegraphics[width=0.23\textwidth]{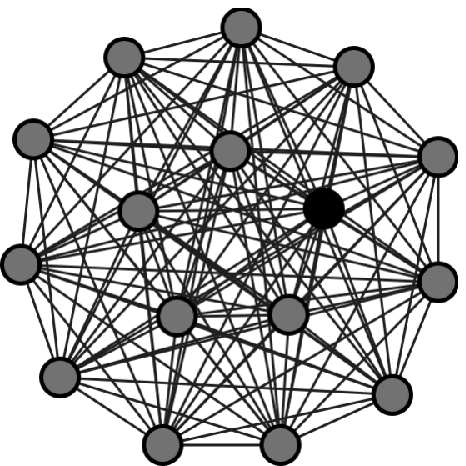}}
\end{subfigure}
\hfill
\begin{subfigure}[$T_{PT}=281$ ]{\label{fnetpt281}
\includegraphics[width=0.23\textwidth]{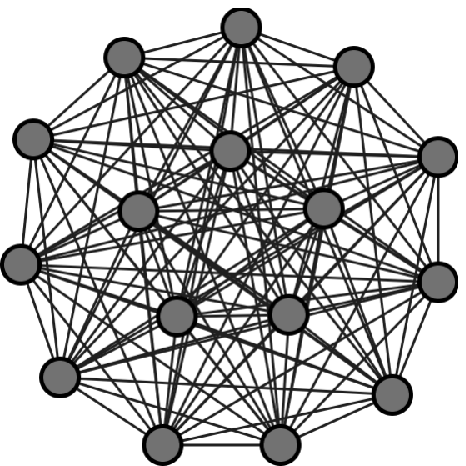}}
\end{subfigure}
\hfill
\begin{subfigure}[$T_{USA}=0$]{\label{fnetusa0}
\includegraphics[width=0.23\textwidth]{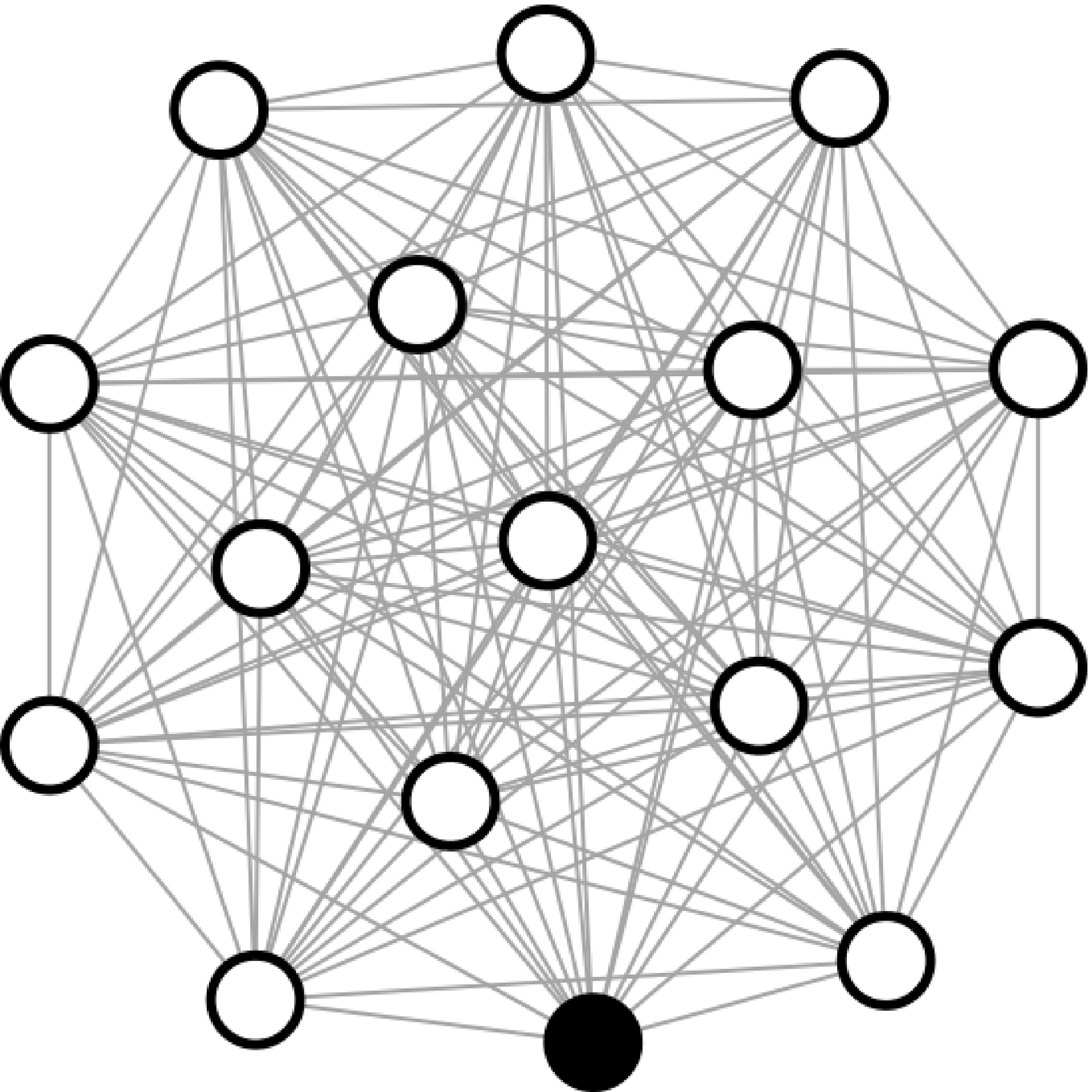}}
\end{subfigure}
\hfill
\begin{subfigure}[$T_{USA}=1$]{\label{fnetusa1}
\includegraphics[width=0.23\textwidth]{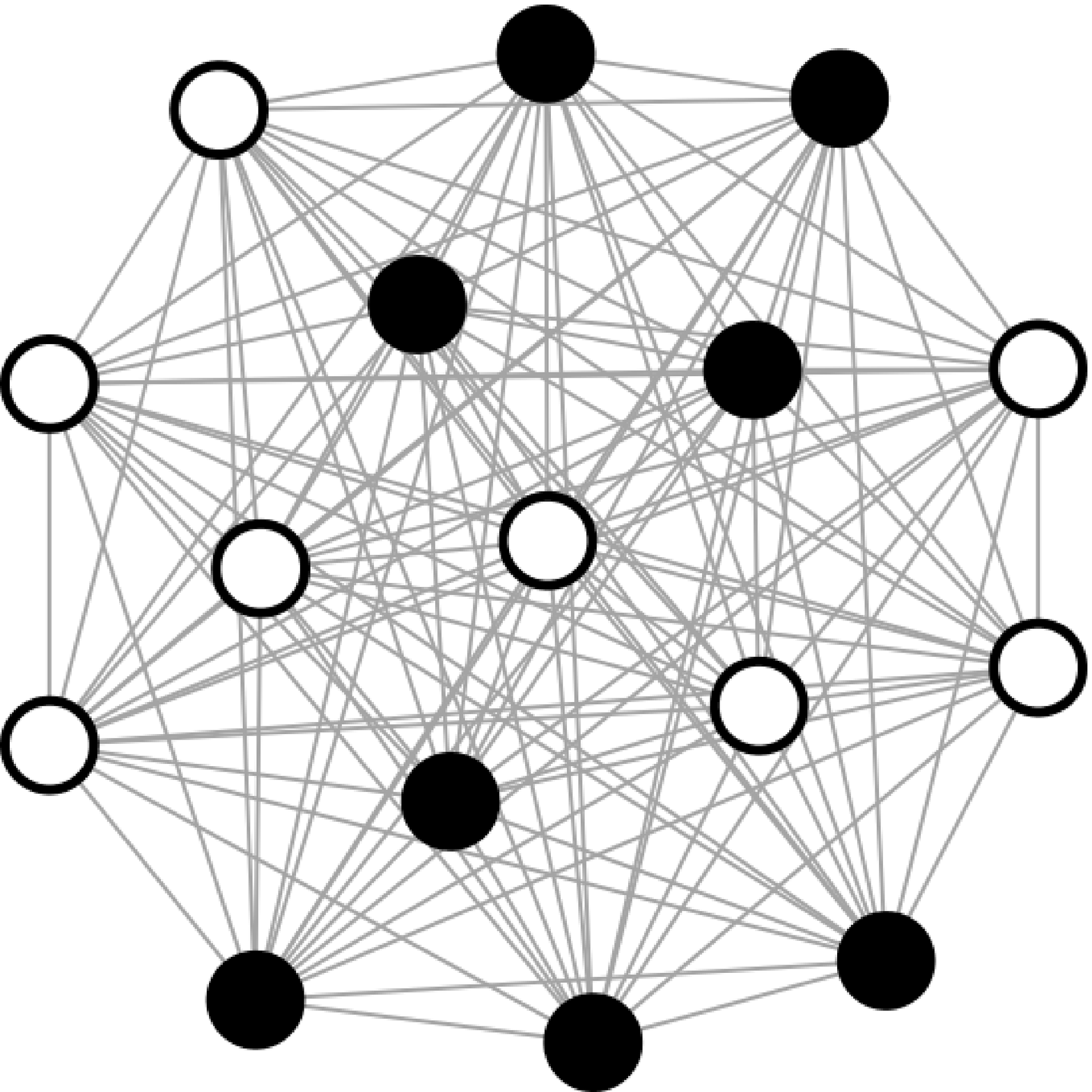}}
\end{subfigure}
\vfill
\begin{subfigure}[$T_{USA}=2$]{\label{fnetusa2}
\includegraphics[width=0.23\textwidth]{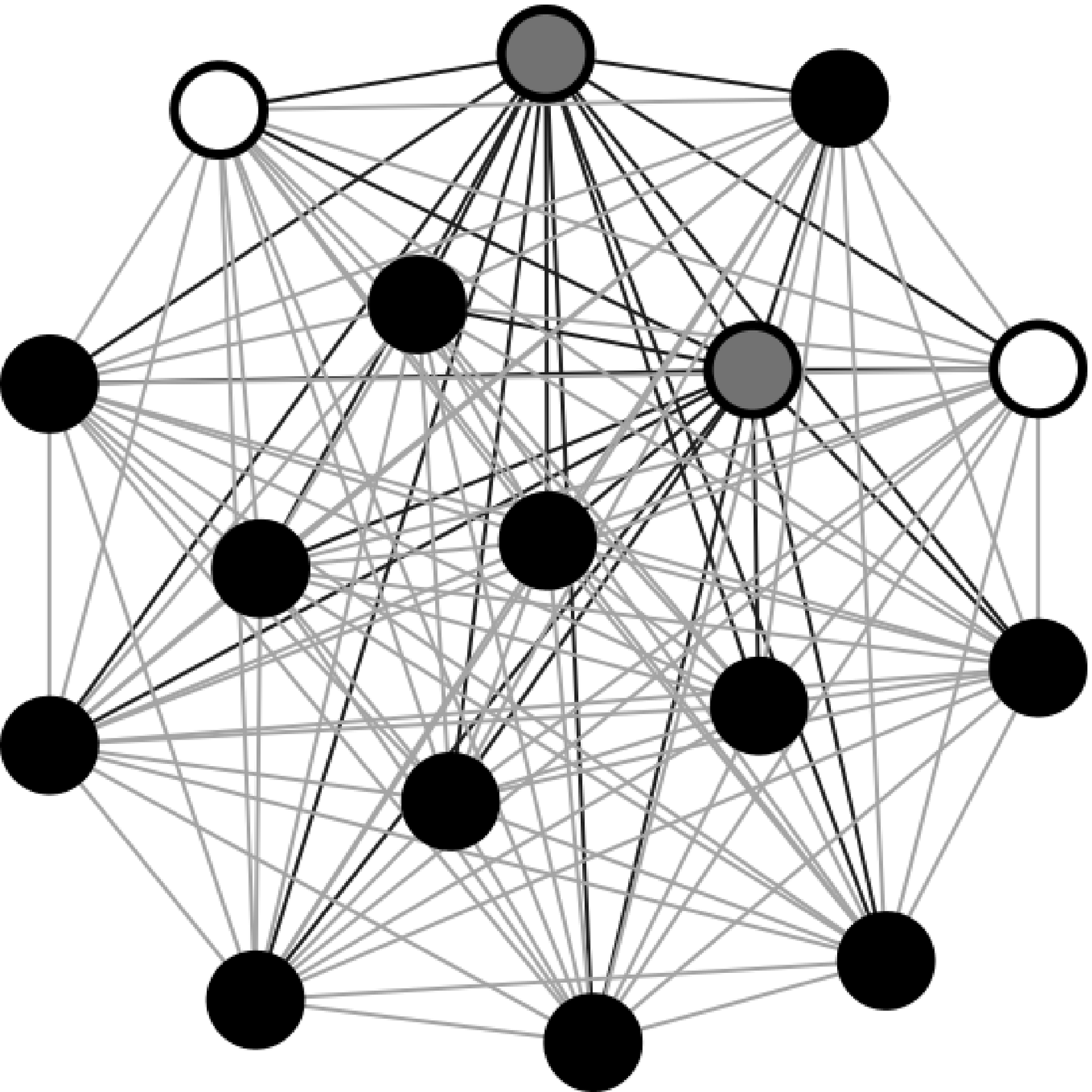}}
\end{subfigure}
\hfill
\begin{subfigure}[$T_{USA}=5$]{\label{fnetusa5}
\includegraphics[width=0.23\textwidth]{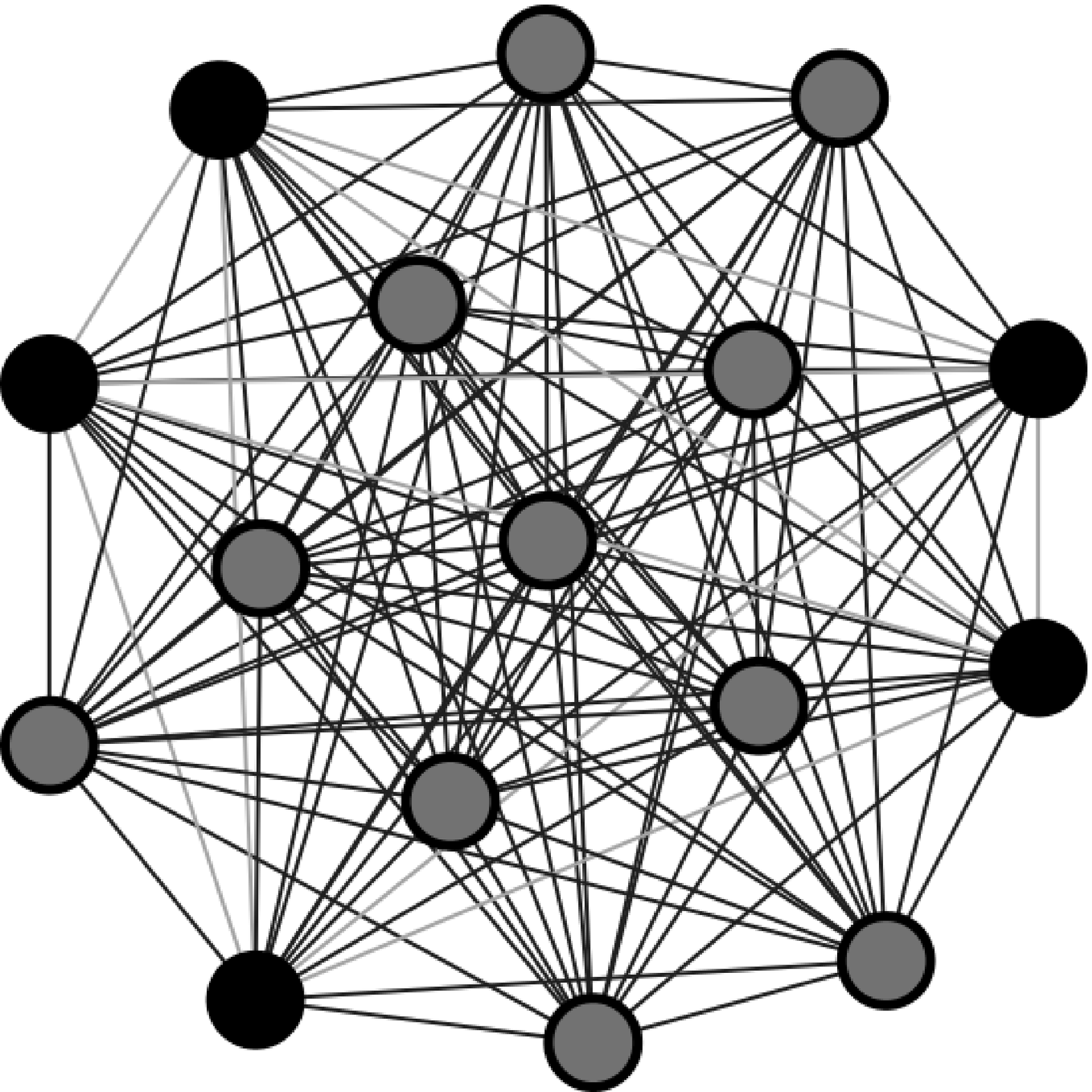}}
\end{subfigure}
\hfill
\begin{subfigure}[$T_{USA}=14$]{\label{fnetusa14}
\includegraphics[width=0.23\textwidth]{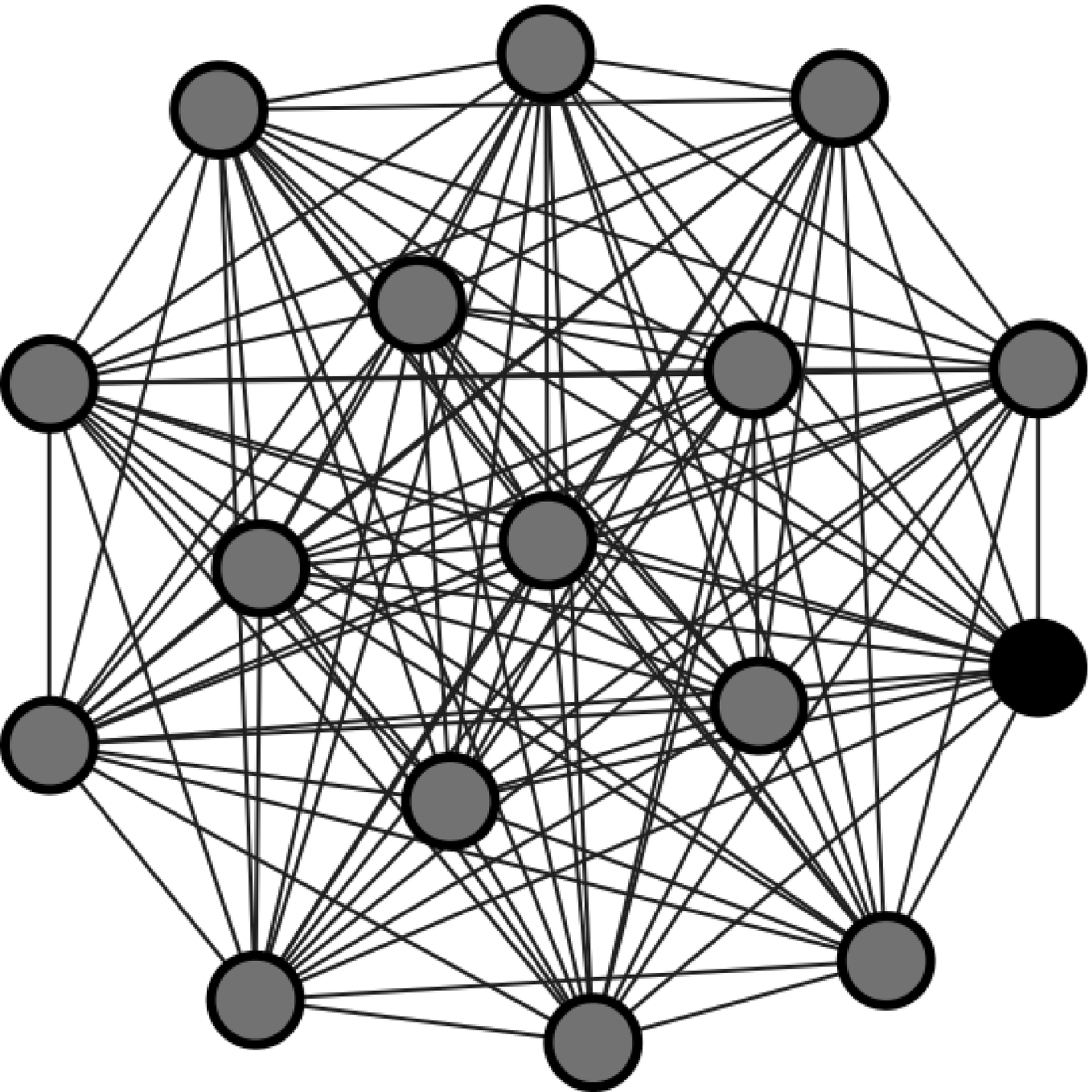}}
\end{subfigure}
\hfill
\begin{subfigure}[$T_{USA}=15$]{\label{fnetusa15}
\includegraphics[width=0.23\textwidth]{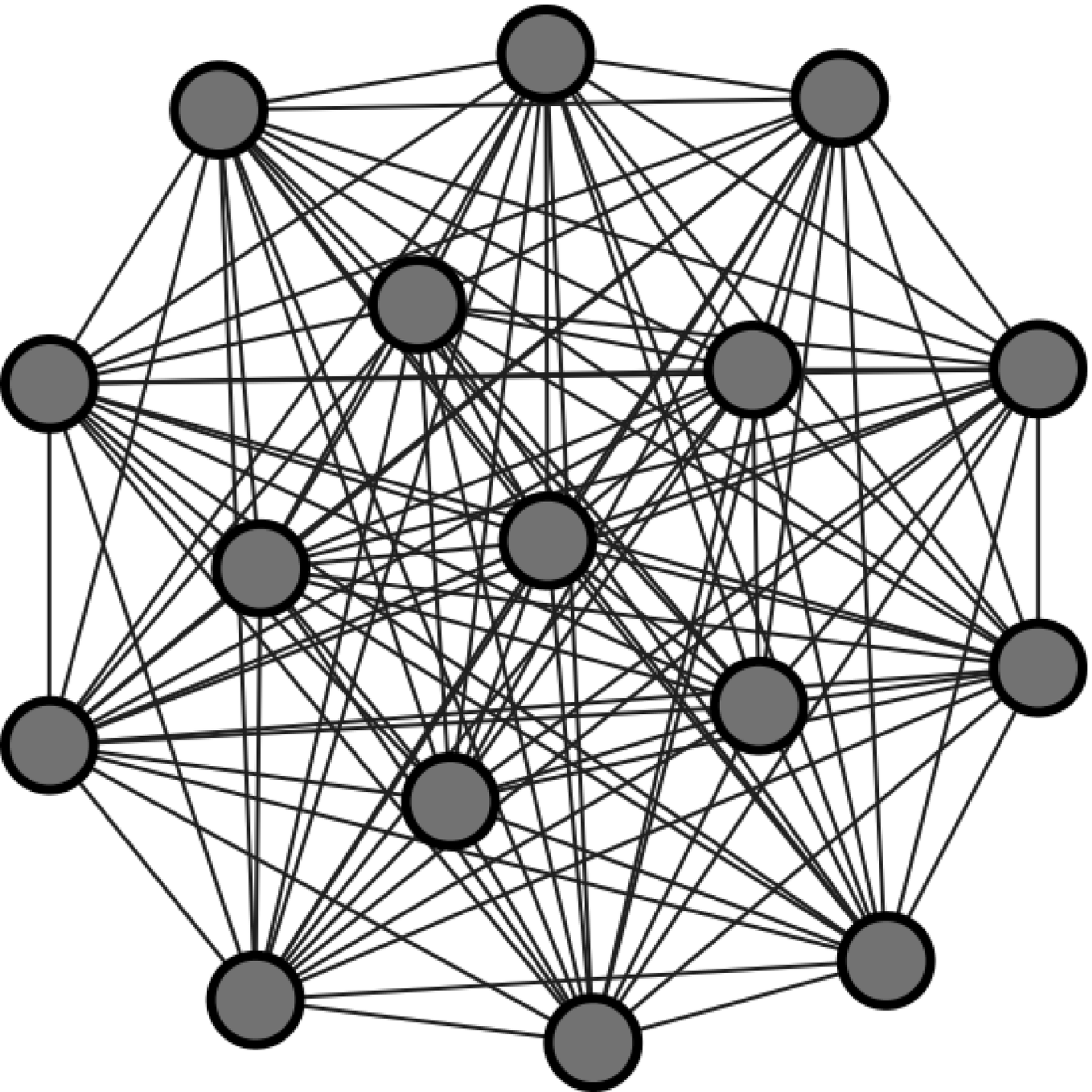}}
\end{subfigure}
\vfill
\begin{subfigure}[$T_{CH}=0$]{\label{fnetch0}
\includegraphics[width=0.23\textwidth]{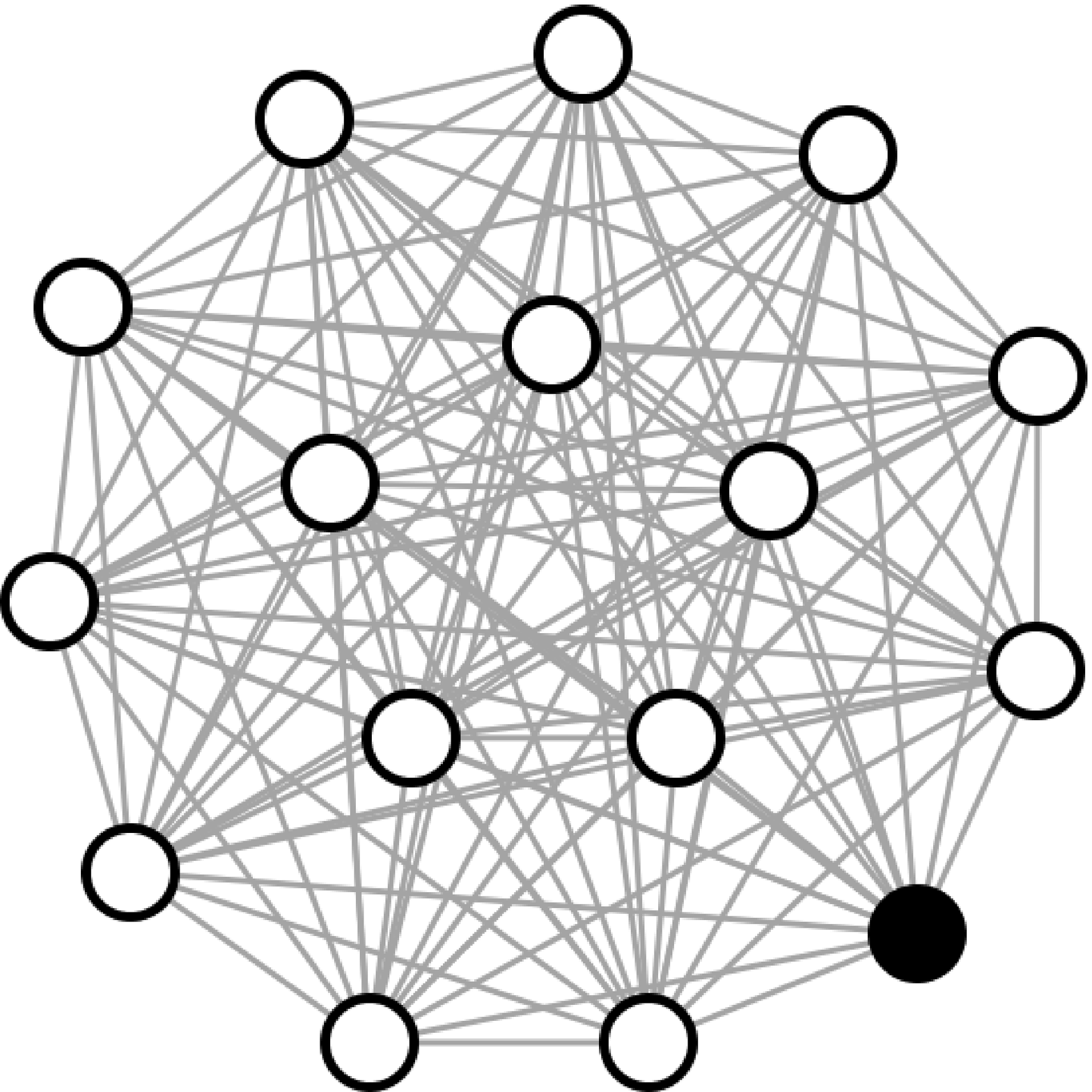}}
\end{subfigure}
\begin{subfigure}[$T_{CH}=1$]{\label{fnetch1}
\includegraphics[width=0.23\textwidth]{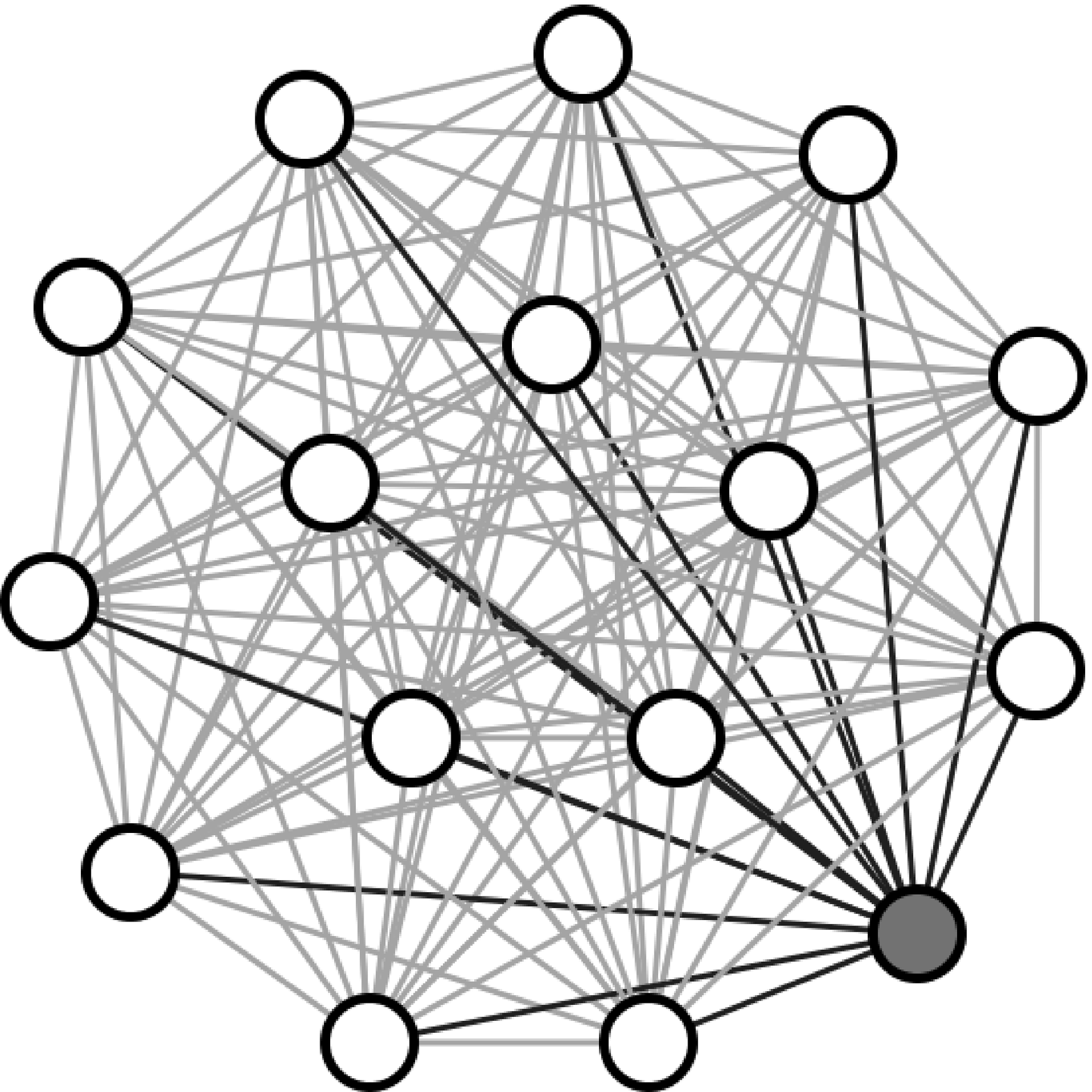}}
\end{subfigure}
\caption{Virus spreading in the network of countries with parameters 
$\beta$ and $\gamma$ taken from Figure~\ref{fdata}; 
\eqref{fnetpt0}--\eqref{fnetpt281} -- initially infected country is Portugal (PT); 
\eqref{fnetusa0}--\eqref{fnetusa15} -- initially infected country is United States (USA); 
\eqref{fnetch0}--\eqref{fnetch1} -- initially infected country is Switzerland (CH). 
Nodes in white mean ``Susceptible''; nodes in black mean ``Infected''; nodes in grey mean ``Recovered''.}
\label{fnet}
\end{figure}

Figure~\ref{fnet} demonstrates the spread of the financial virus through the network, 
where each node represents a random country from the considered list represented 
in Figure~\ref{fdata}. At the initial moment, all nodes are susceptible (white colour) 
except one infected node (black colour). In each time step (``days''), the ``nodes'' 
check whether they have an infection, and an infected node attempts 
to infect all of its neighbours. ``Days'' is an arbitrary unit  
during which the ``nodes'' check and change their status.
If an infection has been detected, then there is a probability 
of $\beta$ that susceptible neighbours will get an infection and change 
their colour to black, and there is a probability $\gamma$ that an infection 
will be removed and the nodes will be recovered. 
Recovered nodes (grey colour) cannot be infected.
When a node becomes recovered, the links between it and its neighbours are darkened, 
since they are no longer possible vectors for contagion spreading \cite{netlogprog}. 
It is important to notice that, in reality, the country's financial system cannot recover evermore. 
Therefore, applying this model, we consider that recovered countries are resistant for some 
short period of time, and then they again become susceptible to the virus.
For the case where the infection begins from Portugal 
(Figures~\ref{fnetpt0}--\ref{fnetpt281}), 
the first infected node (Figure~\ref{fnetpt0}) spreads 
the virus to one of its neighbours at time $T=9$ (Figure~\ref{fnetpt9}). 
At time $T=15$ (Figure~\ref{fnetpt15}) the contagion process slowly continues,
and the last infected node (Figure~\ref{fnetpt280}) changes its state 
to recovered at time $T=281$ (Figure~\ref{fnetpt281}).

It is easy to see that the chain reaction of infection and recovery of nodes 
occurs much faster when it starts from United States of America 
(Figures~\ref{fnetusa0}--\ref{fnetusa15}).
The initially infected node (Figure~\ref{fnetusa0}) spreads 
the infection to neighbouring nodes in the next time step $T=1$ (Figure~\ref{fnetusa1}).
The maximum number of infected nodes is reached at time $T=2$ (Figure~\ref{fnetusa2}).
At time $T=5$ (Figure~\ref{fnetusa5}) most of the infected nodes have already been recovered
and the last infected node (Figure~\ref{fnetusa14}) changed its state to recovered 
at time $T=15$ (Figure~\ref{fnetusa15}).

In the case when Switzerland is initially infected, the virus is not transmitted 
to the neighbours and the infected node is immediately recovered 
(Figures~\ref{fnetch0}--\ref{fnetch1}).


\subsection{Epidemiological SIR model}

\noindent The initial value problem \eqref{eqSIR}--\eqref{eqIC} 
can be solved using a numerical approach. In practice, the solution 
can be obtained in the form of a time-series function of each compartment.
In our work we solve the system of differential equations in MATLAB.
The obtained results are consistent with those that were obtained 
with the network simulations.

\begin{figure}[ht!]
\centering
\begin{subfigure}[PT: Portugal]{\label{fsirpt}
\includegraphics[width=0.32\textwidth]{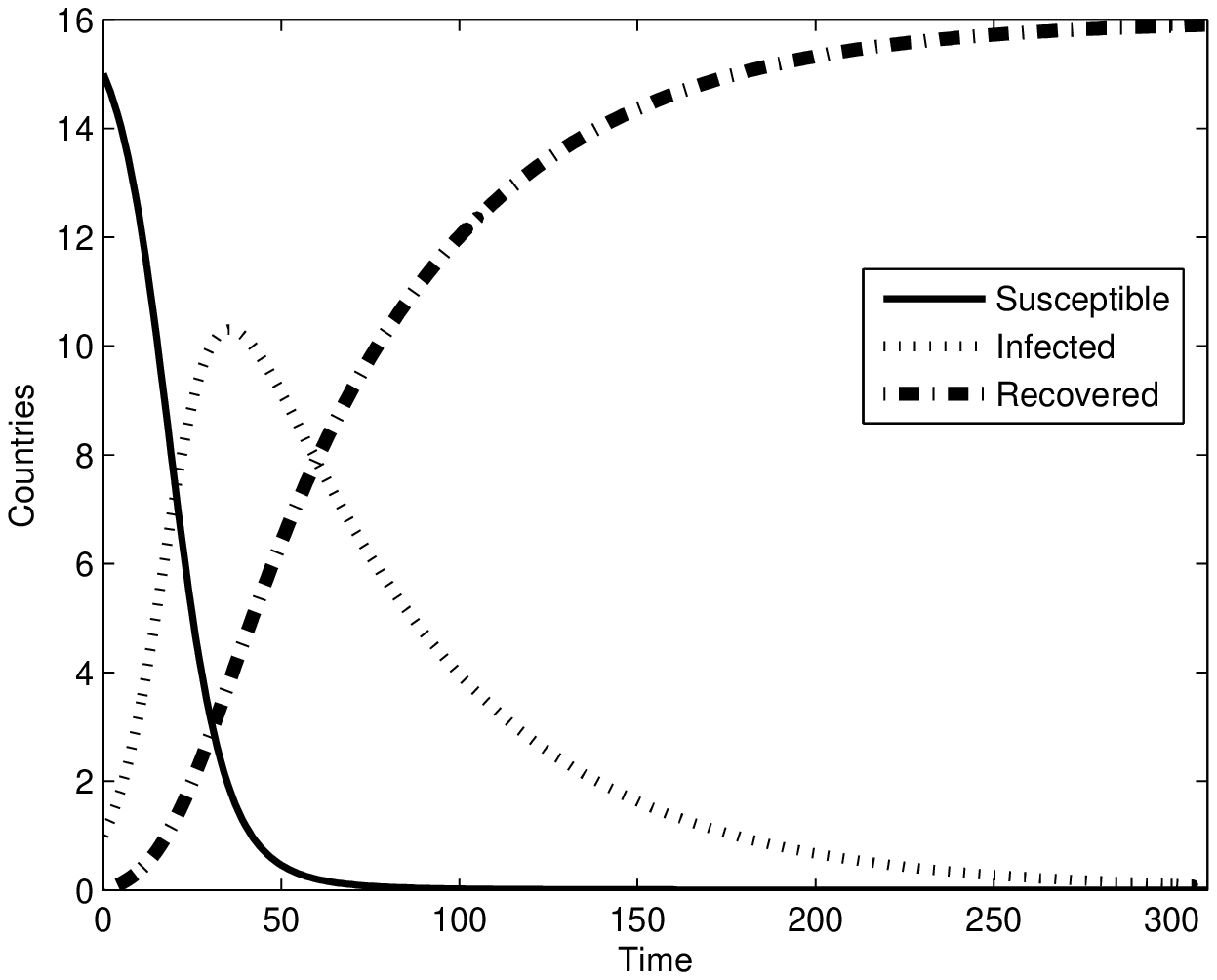}}
\end{subfigure}
\begin{subfigure}[USA: United States]{\label{fsirusa}
\includegraphics[width=0.32\textwidth]{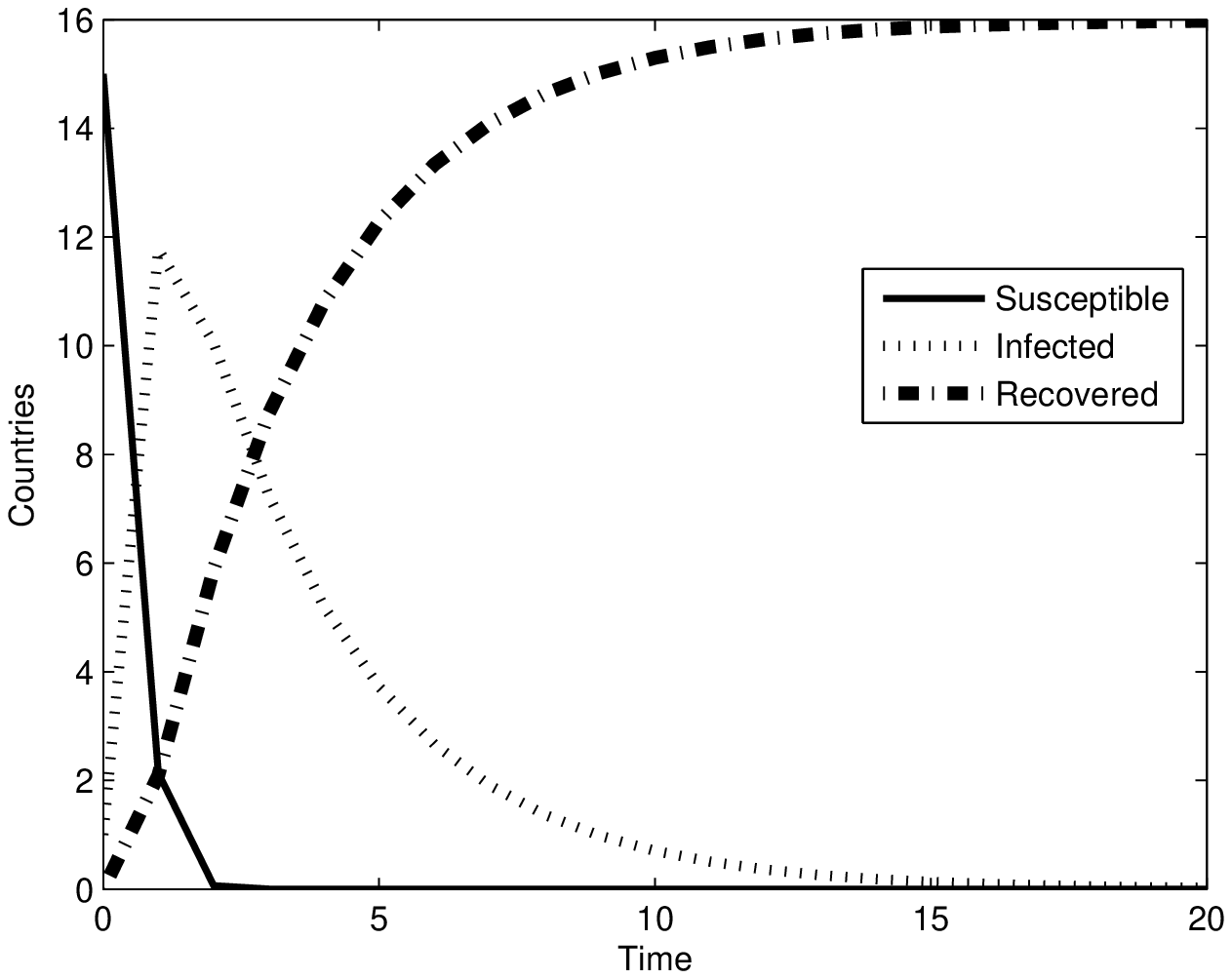}}
\end{subfigure}
\begin{subfigure}[CH: Switzerland]{\label{fsirch}
\includegraphics[width=0.32\textwidth]{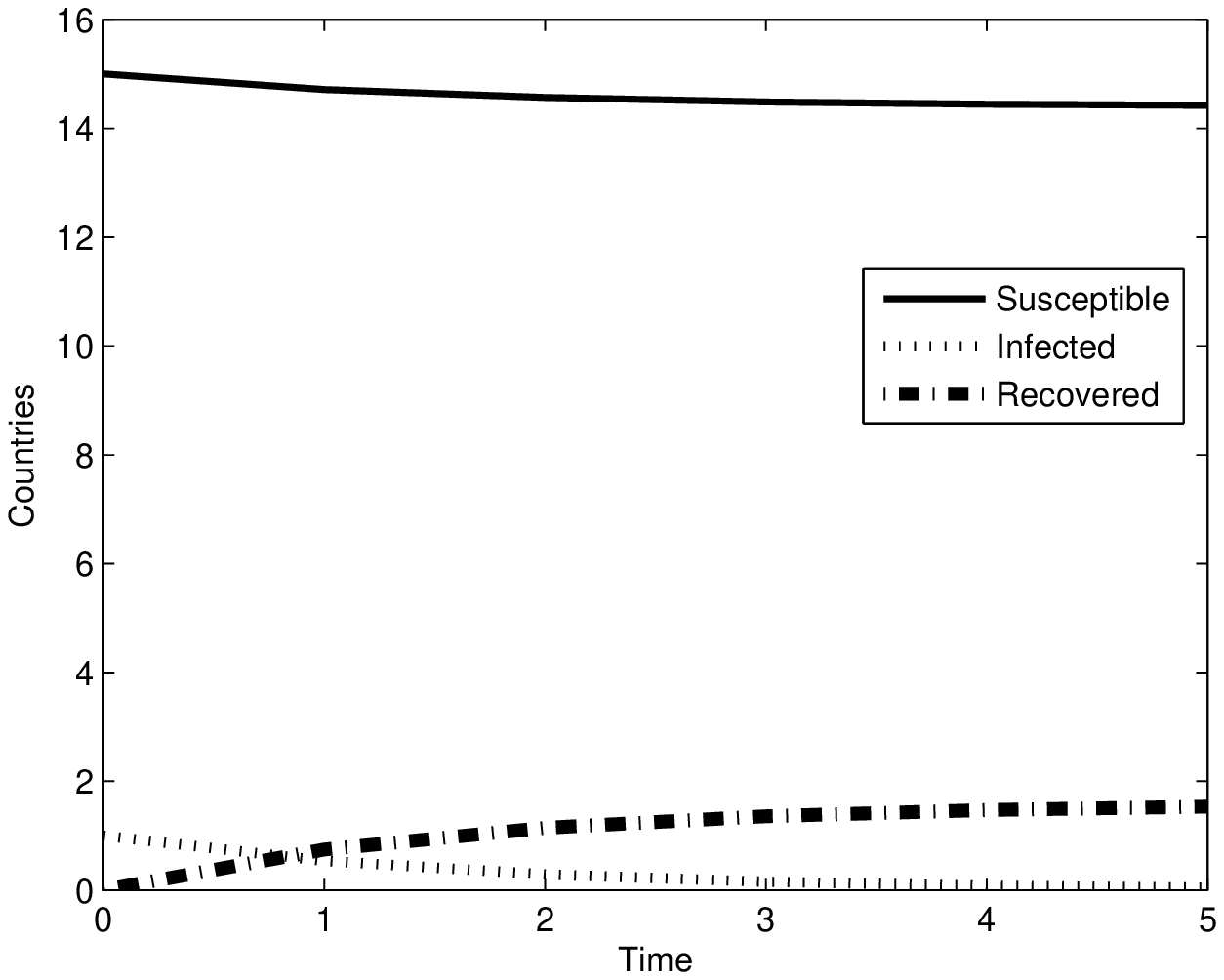}}
\end{subfigure}
\caption{The SIR contagion risk model \eqref{eqSIR}--\eqref{eqIC} with 
parameters $\beta$ and $\gamma$	for Portugal, United States, and Switzerland, 
taken from Figure~\ref{fdata}; the initial conditions are $S(0)=15$, $I(0)=1$, $R(0)=0$.}
\label{fsir}
\end{figure}

The behaviour of the epidemiological model for Portugal, 
United States of America, and Switzerland parameters,
are shown in Figure~\ref{fsir}.
When infection spreading begins from Portugal (Figure~\ref{fsirpt}), 
contagion has almost reached the contagion-free equilibrium ($I(T)=0$) 
after 281 time steps. The spread of contagion occurs over a long period 
of time and the recovery process goes slowly too.
If the United States is the starting point for virus spreading (Figure~\ref{fsirusa}),
the contagion spreads rapidly and affects a large number of countries 
in a short period of time, and then swiftly decreases as the recovery process takes fast.
If the initially infected country is Switzerland, the virus immediately dies out (Figure~\ref{fsirch}).

The results in Figure~\ref{fsir} coincide with those that were obtained earlier in Figure~\ref{fnet}.
It means that both methods of modelling of contagion spreading are in agreement with each other.

\begin{figure}[hp]
\centering
\begin{subfigure}[BE: Belgium]{\label{fsircbe}
\includegraphics[scale=0.30]{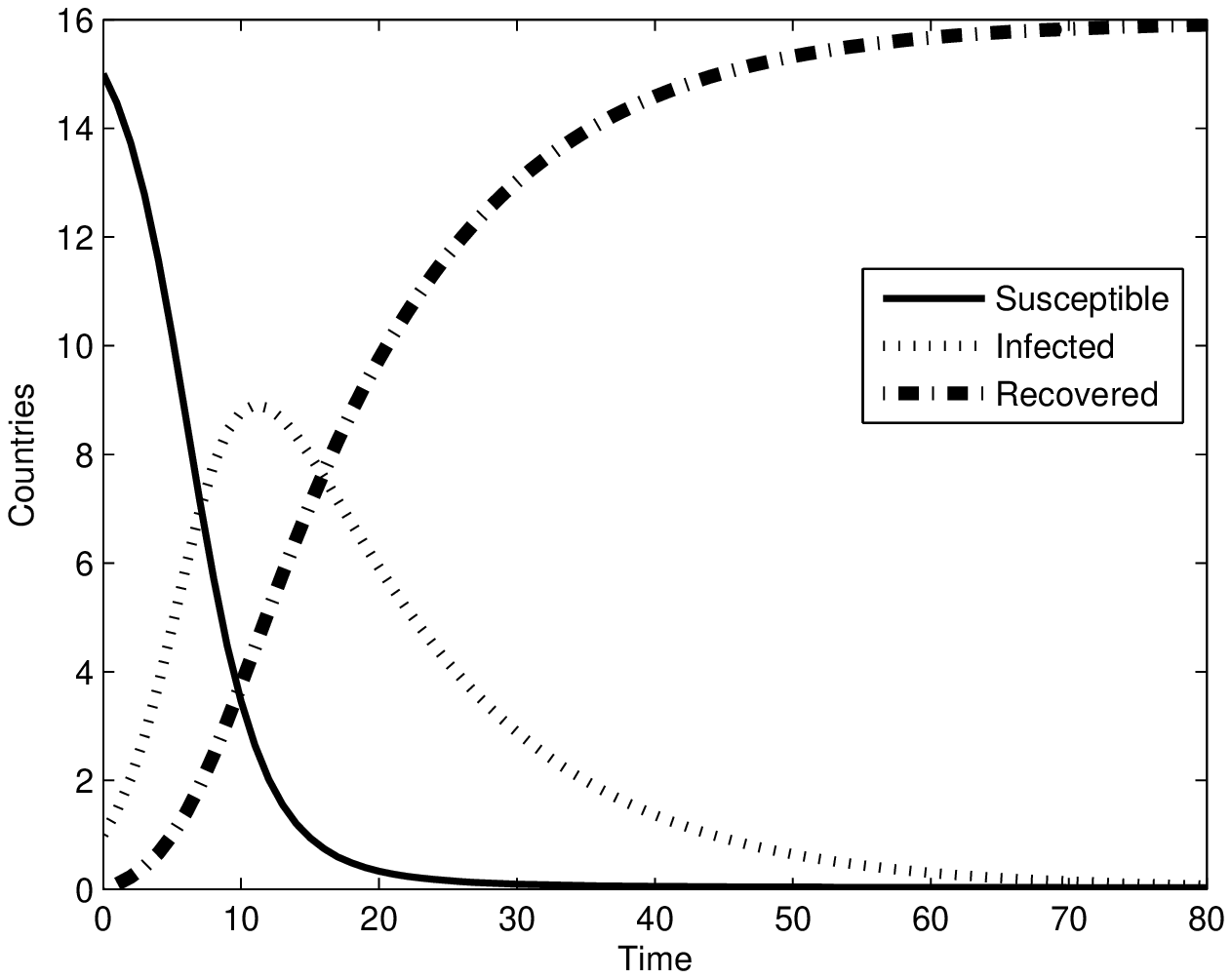}}
\end{subfigure}
\begin{subfigure}[ES: Spain]{\label{fsirces}
\includegraphics[scale=0.30]{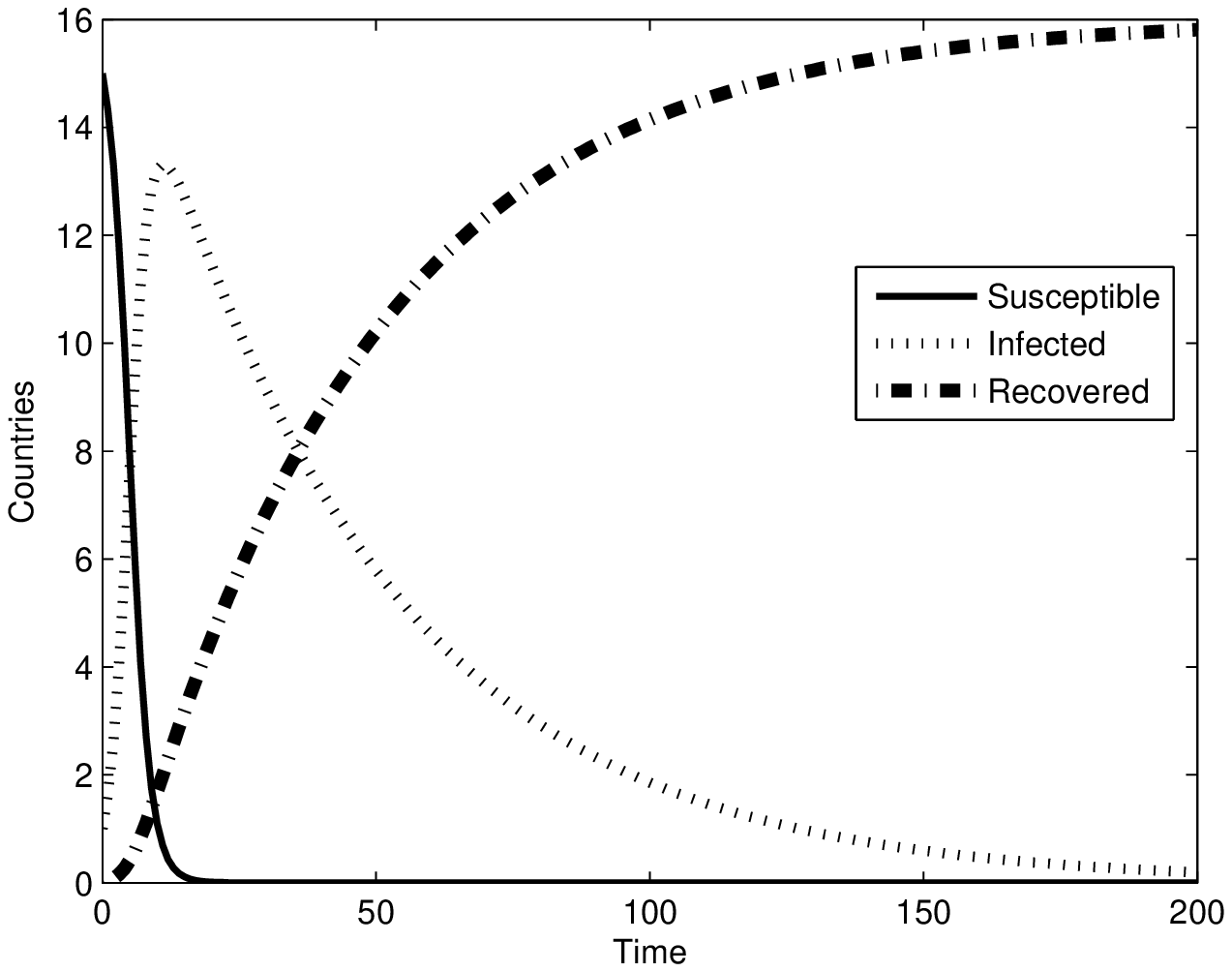}}
\end{subfigure}
\begin{subfigure}[GR: Greece]{\label{fsircgr}
\includegraphics[scale=0.30]{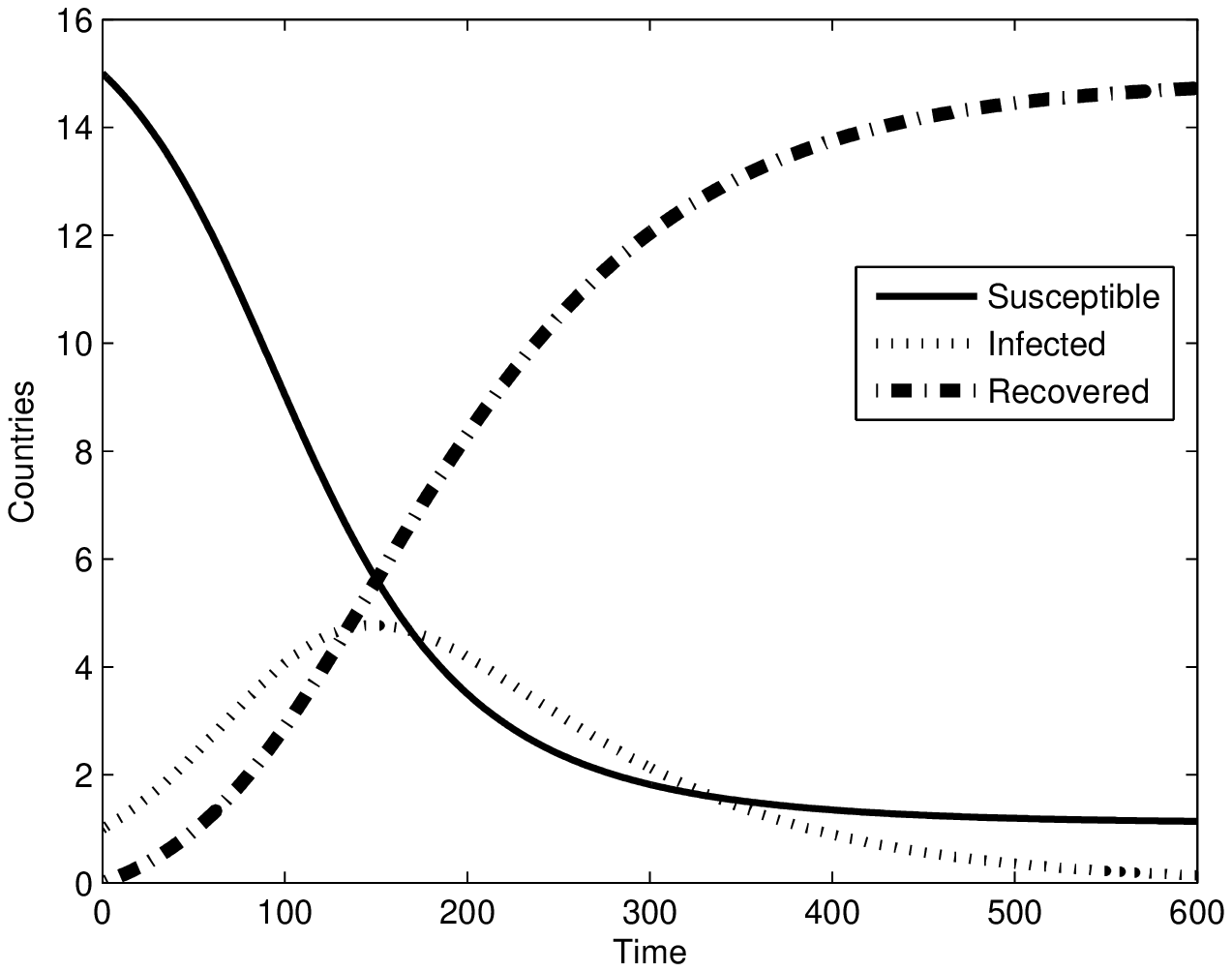}}
\end{subfigure}
\begin{subfigure}[IE: Ireland]{\label{fsircie}
\includegraphics[scale=0.30]{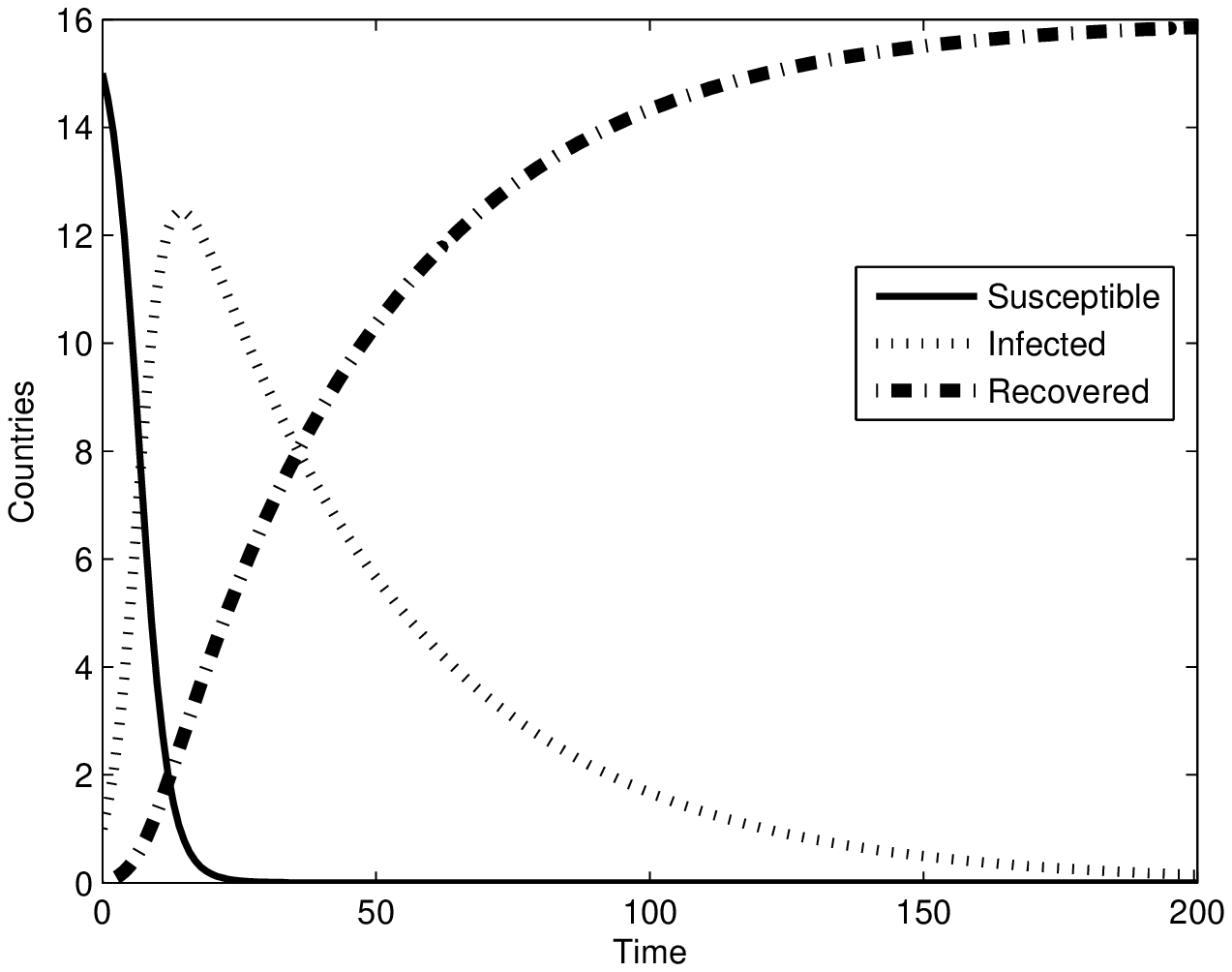}}
\end{subfigure}
\begin{subfigure}[IT: Italy]{\label{fsircit}
\includegraphics[scale=0.30]{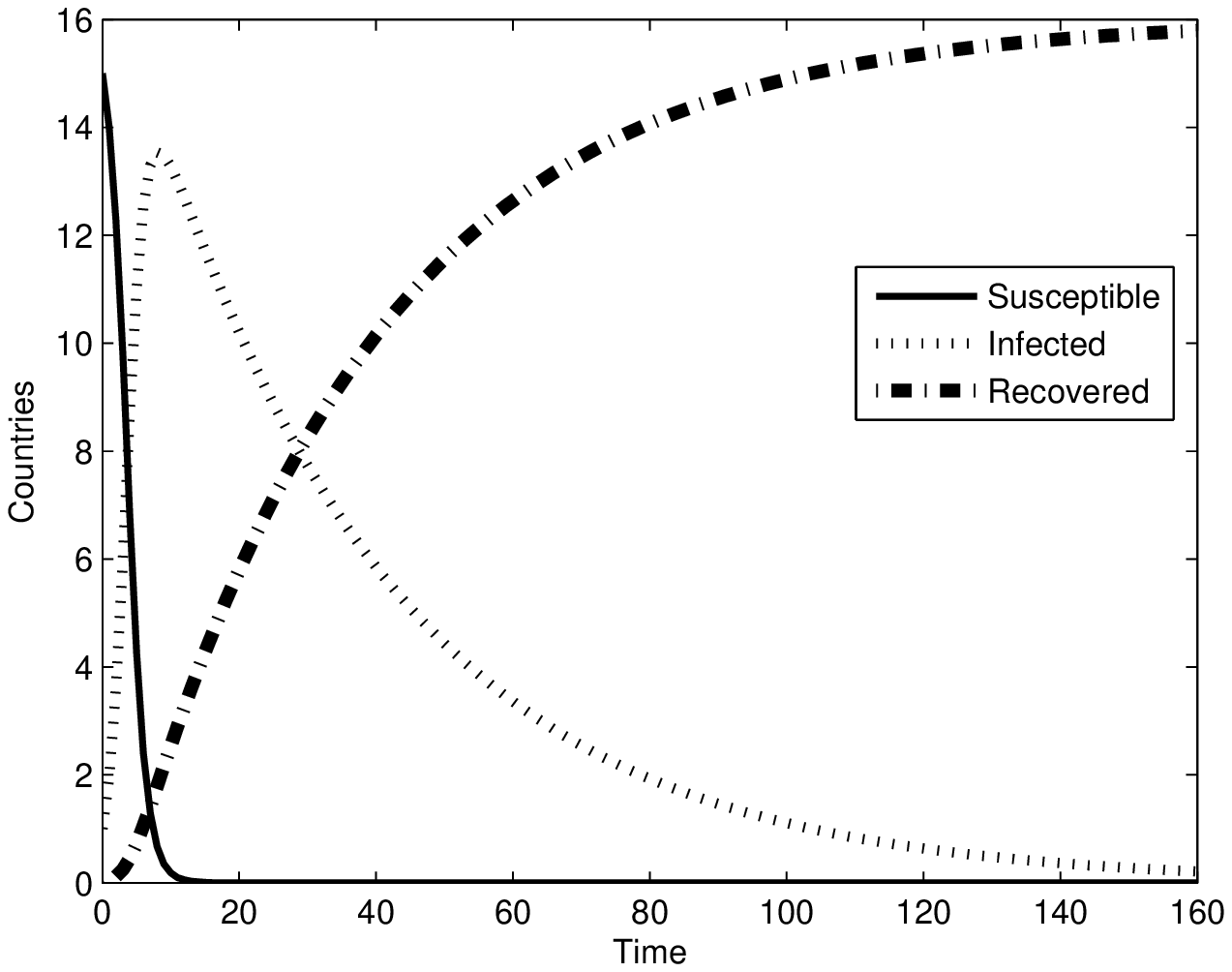}}
\end{subfigure}
\begin{subfigure}[JP: Japan]{\label{fsircjp}
\includegraphics[scale=0.30]{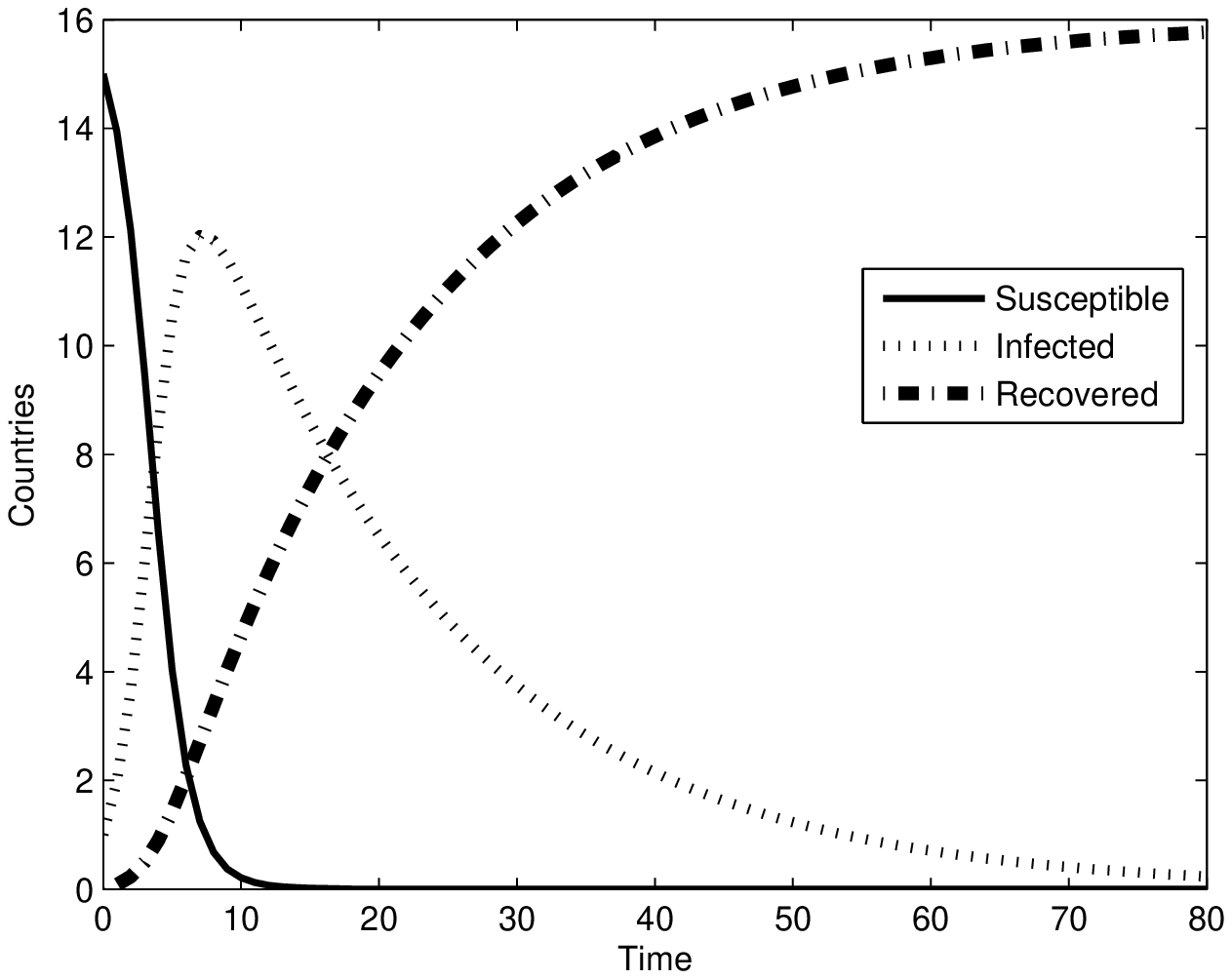}}
\end{subfigure}
\begin{subfigure}[AT: Austria]{\label{fsircat}
\includegraphics[scale=0.30]{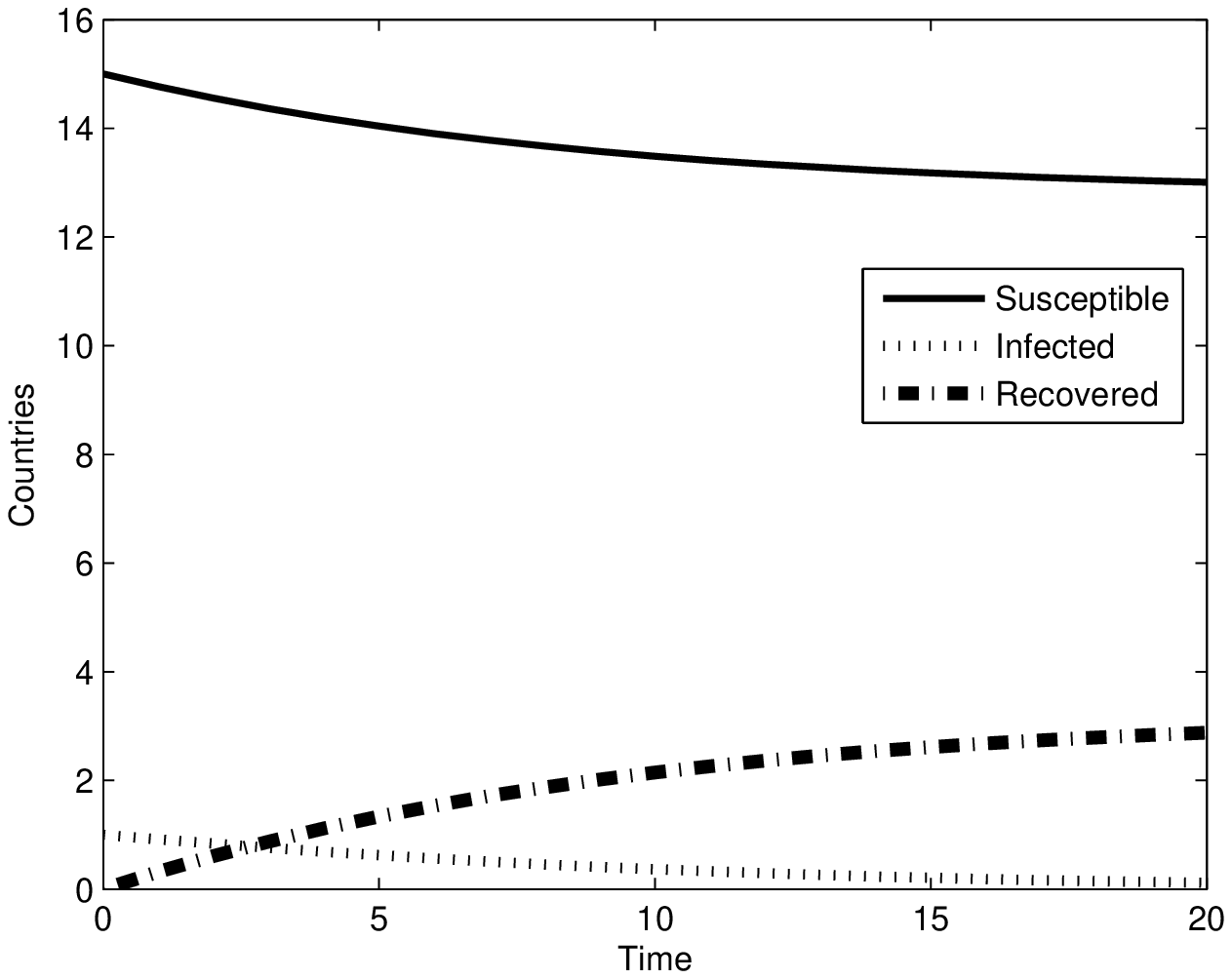}}
\end{subfigure}
\begin{subfigure}[FR: France]{\label{fsircfr}
\includegraphics[scale=0.30]{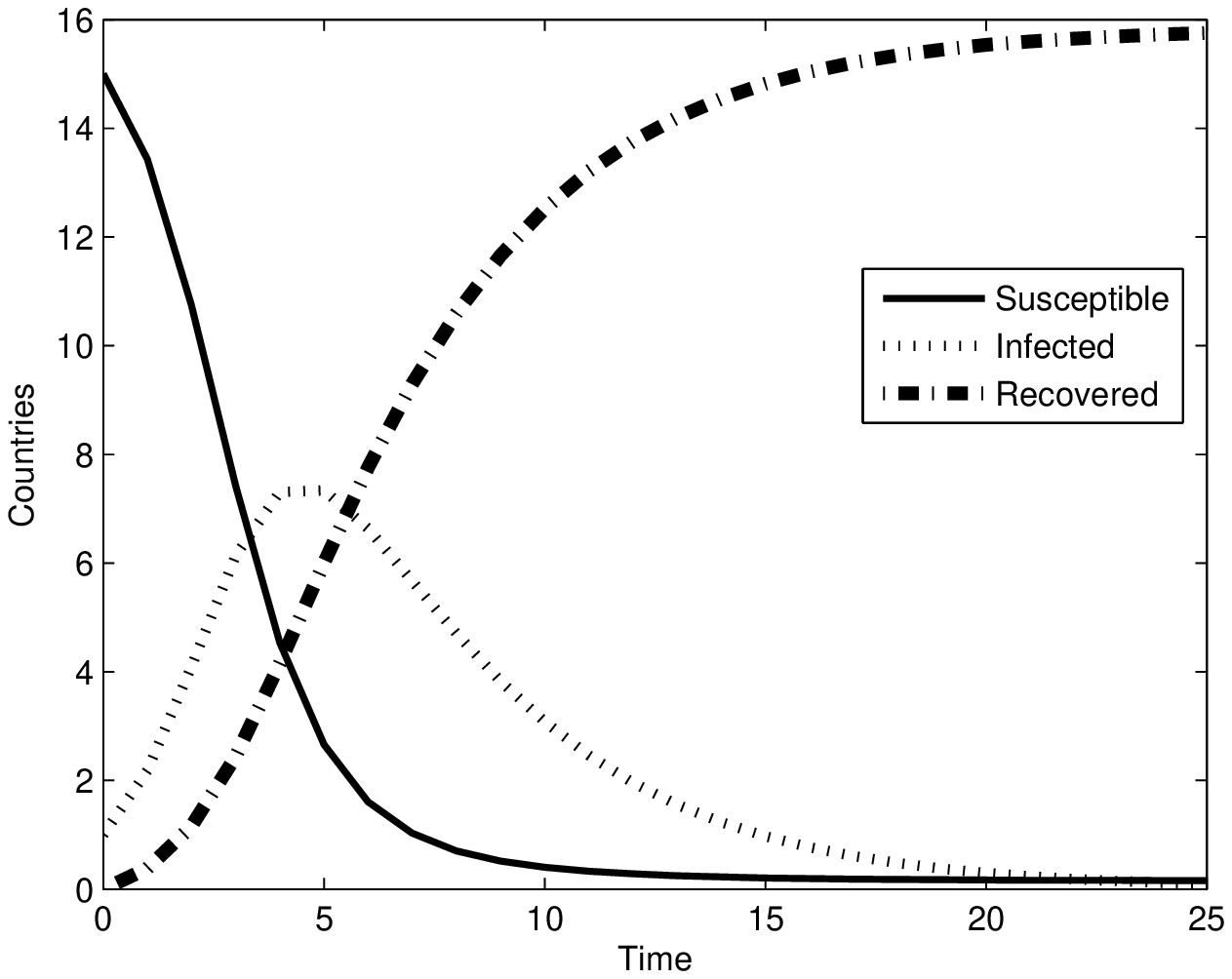}}
\end{subfigure}
\begin{subfigure}[DE: Germany]{\label{fsircde}
\includegraphics[scale=0.30]{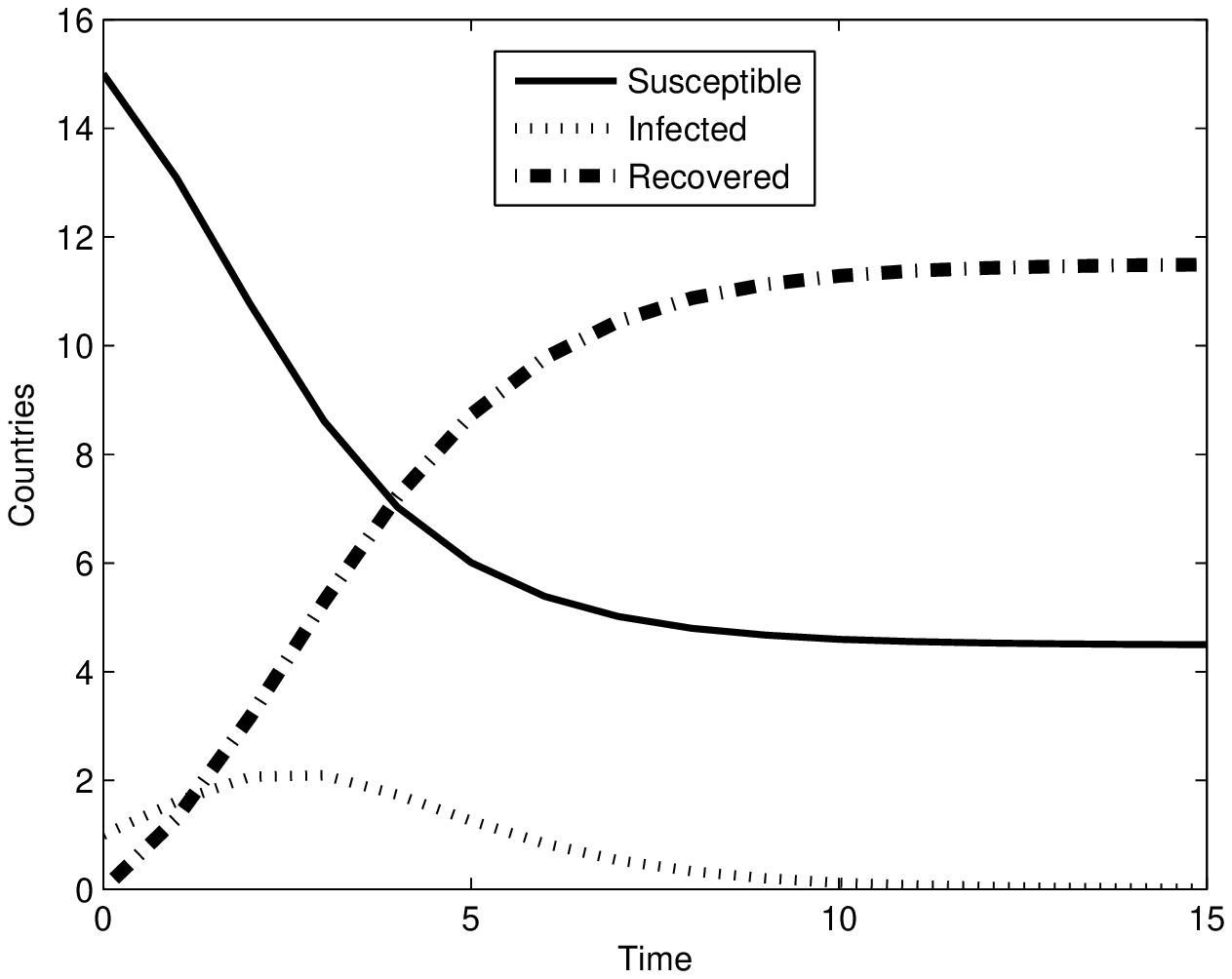}}
\end{subfigure}
\begin{subfigure}[GB: United Kingdom]{\label{fsircgb}
\includegraphics[scale=0.30]{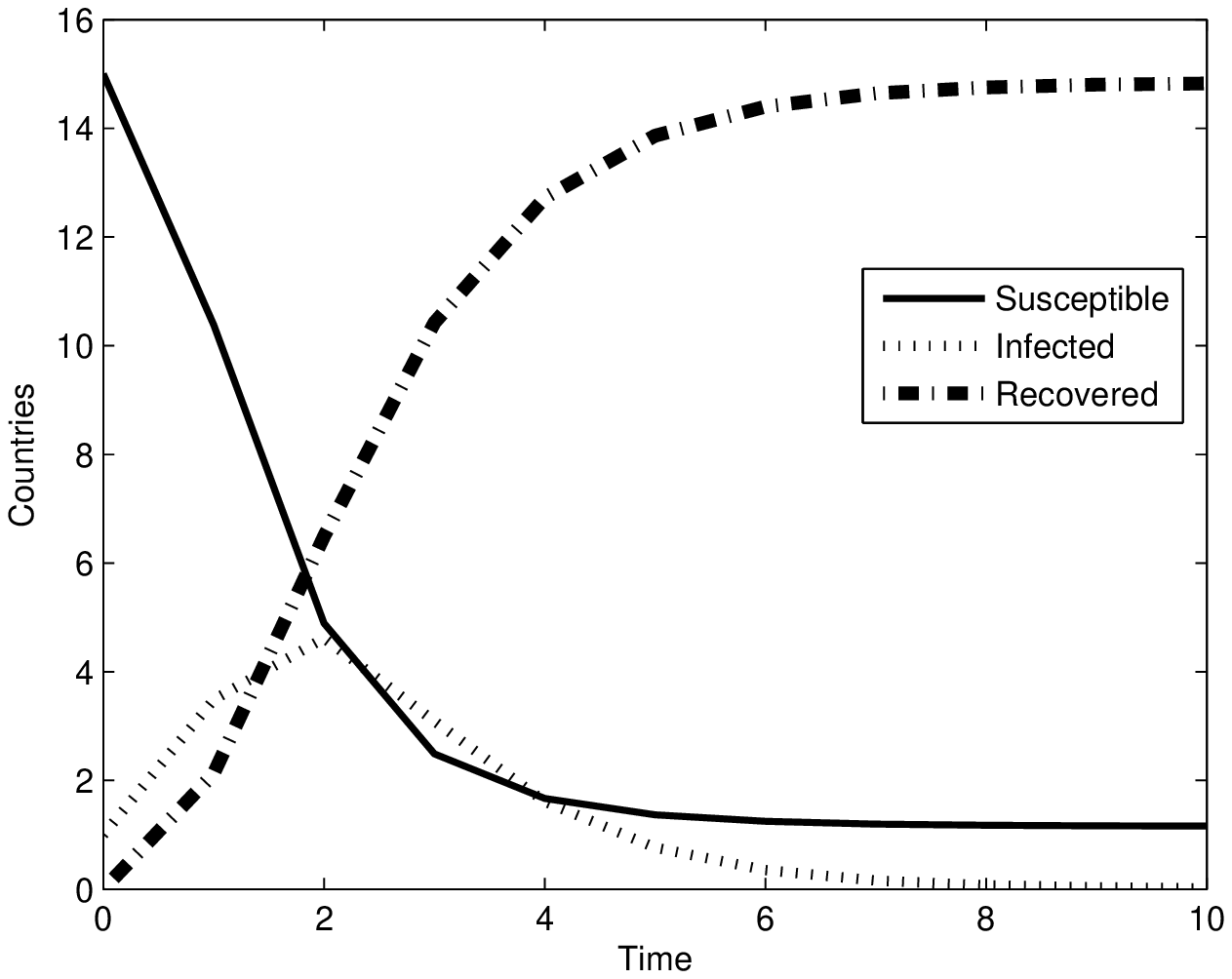}}
\end{subfigure}
\begin{subfigure}[NL: Netherlands]{\label{fsircnl}
\includegraphics[scale=0.30]{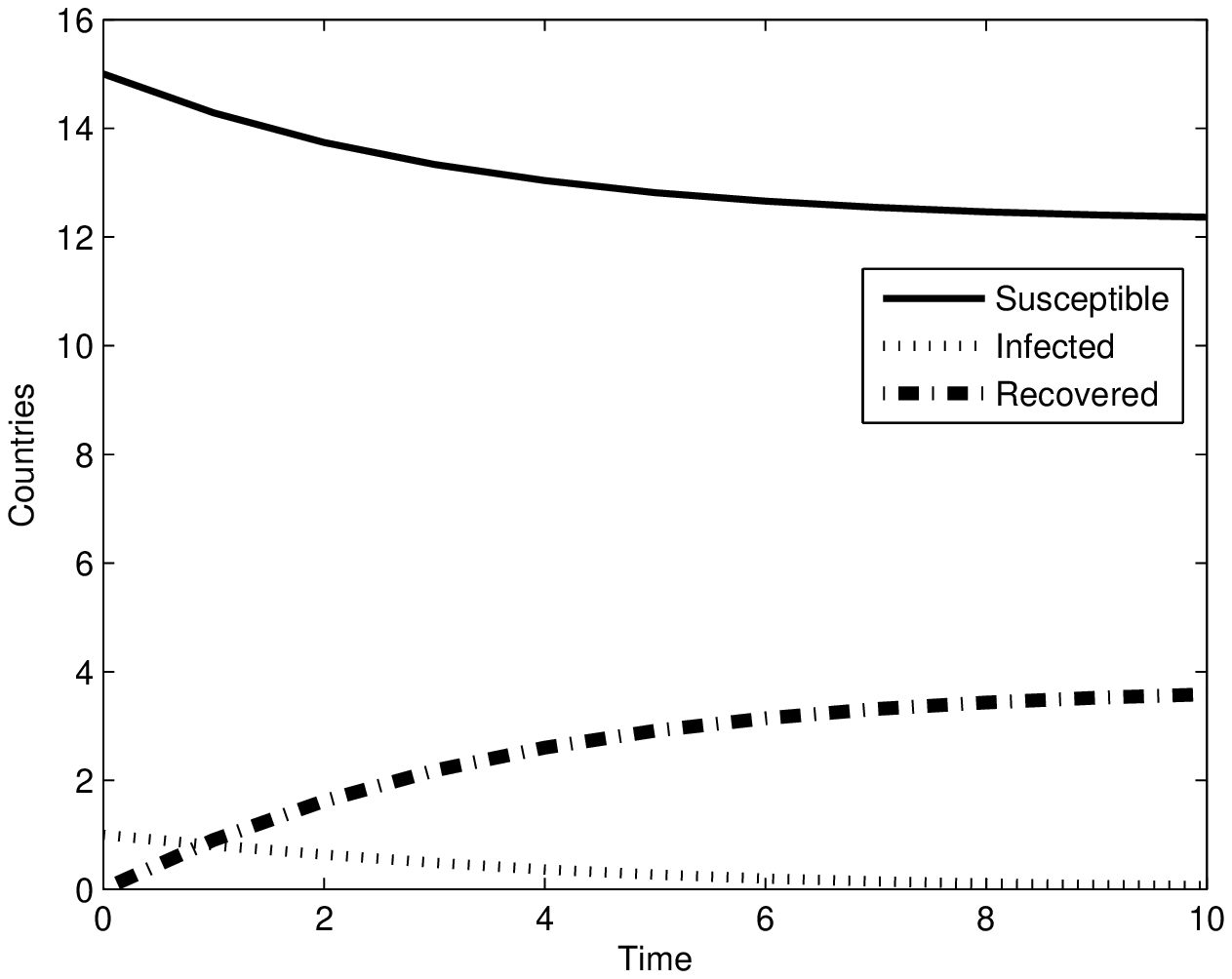}}
\end{subfigure}
\begin{subfigure}[AU: Australia]{\label{fsircau}
\includegraphics[scale=0.30]{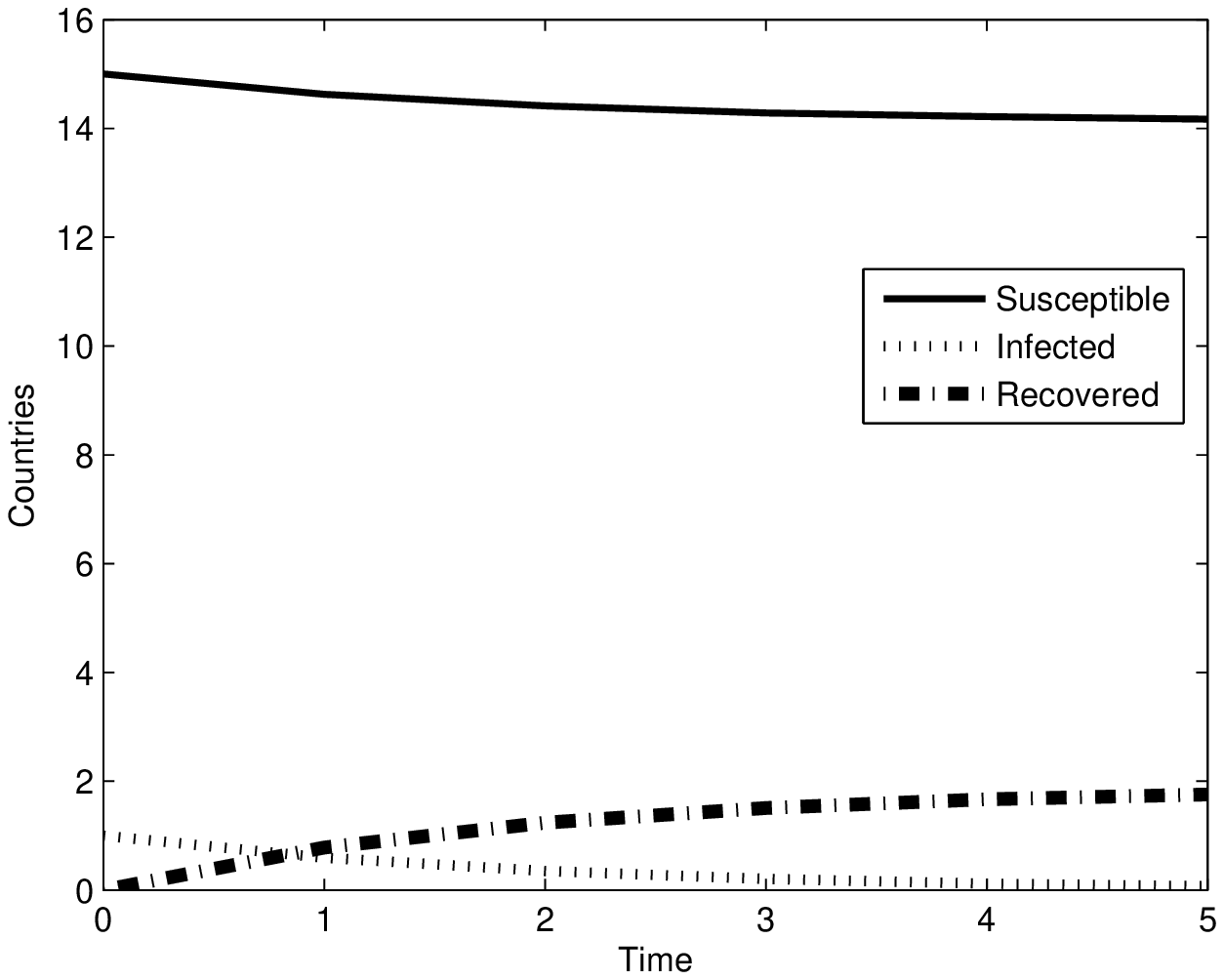}}
\end{subfigure}
\begin{subfigure}[SE: Sweden]{\label{fsircse}
\includegraphics[scale=0.30]{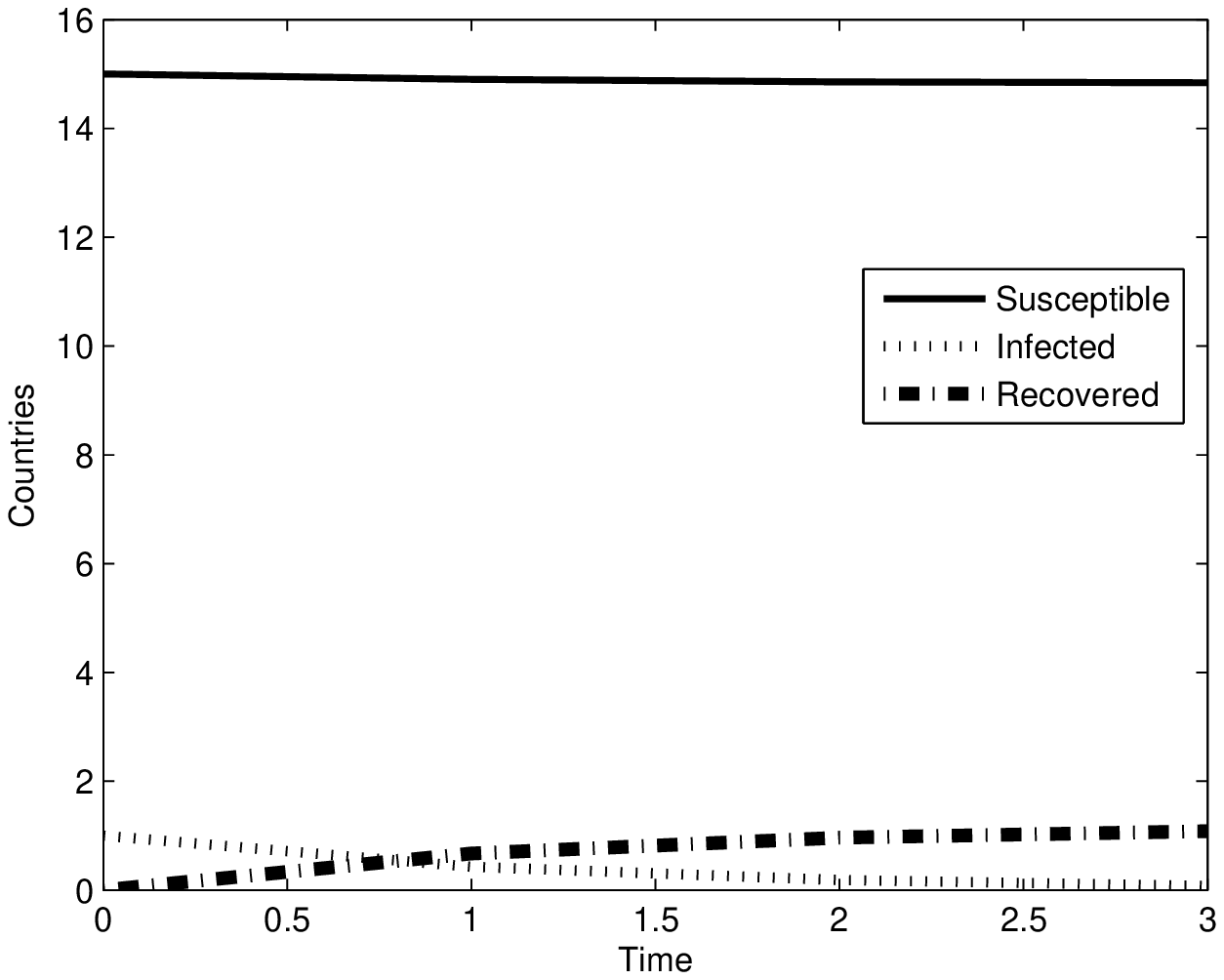}}
\end{subfigure}
\caption{The SIR contagion risk model with parameters $\beta$ 
and $\gamma$ taken from Figure~\ref{fdata}.}
\label{fsirc}
\end{figure}

Figures~\ref{fnet}--\ref{fsir} show that the contagion spreading processes take place 
in different ways, depending on the country where it begins.
The countries that are in Group 3 of Table~\ref{tabular:gamma} 
have the highest recovery rate.
Within a short period of time, the infected will recover 
(Figure~\ref{fsirch} and Figure~\ref{fsircde}--~\ref{fsircse}).
If the infection begins from a country listed in Group~2 of 
Table~\ref{tabular:gamma}, then the contagion ceases to spread 
and all infected become recovered after 10 to 25 time steps 
(Figure~\ref{fsirusa} and Figure~\ref{fsircat}--~\ref{fsircfr}).  
The situation is completely opposite for the countries in Group~1 
of Table~\ref{tabular:gamma}. For them, the virus infect the highest number 
of countries and takes much more time, and the recovery process 
is slower too (Figure~\ref{fsirpt} and Figure~\ref{fsircbe}--~\ref{fsircjp}).

The reason for the identified differences lies in the 
different economic state of the country where contagion begins,
especially in the adequacy of country's reserve capital.
If a country has a big reserve capital and, consequently, 
a high credit rating position, then a high recovery rate indicates its ability 
to cover possible risks in the shortest time period.
The situation is completely opposite for countries with low recovery rate.
If any of these countries will be forced to fulfil their obligations, 
it will be difficult for their economies and, therefore, 
the recovery process will take longer.


\section{Conclusions}
\label{sec:04}

\noindent The recent global crisis of 2008 placed 
the economic analysis as one of the most relevant
political and social concerns of the most indebted countries. 
Here we considered some of the western countries 
in these conditions. Precisely, we investigated 
and modelled the process of contagion spreading 
in a global inter-country network, revealing 
the degree of interconnection of national financial systems, 
identifying the potential systemic financial risks and their
effects. Our research was done with real data from the Bank of International
Settlements on the volumes of consolidated foreign claims on ultimate risk bases in
several countries, and data of credit rating from the Guardian Datablog \cite{guardata}. 
The dynamics of infection spreading of a virus in the financial
system on the given network of countries was simulated
with NetLogo, an agent-based programming language, and 
integrated modelling environment, and confirmed by an
epidemiological SIR model. The infection process was shown 
to depend on the parameter value of the recovery rate, 
as well as on the country, which initially begins the process of infection.
We found out that if one of the financially unstable countries will be the starting 
point in the spread of contagion, and will be forced to fulfil its obligations 
as a counterparty, then the global financial system will have serious problems, 
the negative effects of which will continue during a long period of time. 
Therefore, the countries with a powerful economy and good credit rating position 
are more reliable counterparties, since if necessary 
they will be able to fulfil their obligations.

According to the standard SIR methodology, both parameters $\beta$ and $\gamma$ 
are constant for the entire sample. However, in reality, these data are unique 
for each country and depend on the infection force of the affected country 
and the financial stability of the susceptible. Another line of research,
motivated by the fact that the level of exposure and interference between 
countries and financial institutions is not the same, consists
to consider a more realistic representation as a graph with varying edge weights.
These and other issues are under investigation and will be addressed elsewhere.


\section*{Acknowledgements}

\noindent This research was supported by the Portuguese Foundation 
for Science and Technology (FCT -- Funda\c{c}\~{a}o para a Ci\^{e}ncia e a Tecnologia), 
through CIDMA -- Center for Research and Development in Mathematics and Applications, 
within project UID/MAT/04106/2019. Kostylenko is also supported 
by the FCT Ph.D. fellowship PD/BD/114188/2016.
We are very grateful to Professor Yuriy Petrushenko, Doctor of Economics, 
for providing us with a consultation regarding the data used to calculate 
the beta and gamma parameters in our work; and to four anonymous Reviewers,  
for valuable remarks and comments, which significantly contributed 
to the quality of the paper.


\section*{Conflict of interest}

The authors declare that there is no conflicts of interest in this paper.



\end{document}